\documentclass[aps,prl,reprint,nofootinbib,twocolumn,superscriptaddress,showpacs,showkeys,longbibliography]{revtex4-1}

\usepackage{eurosym}
\usepackage{amsmath,amssymb,amstext}
\usepackage{ulem}
\usepackage[usenames,dvipsnames]{color}
\usepackage{graphicx}
\usepackage{braket}
\usepackage{natbib}
\usepackage{comment}
\usepackage{dcolumn}
\usepackage[english]{babel}
\usepackage{wasysym}
\usepackage[colorlinks,bookmarks=false,citecolor=blue,linkcolor=red,urlcolor=blue]{hyperref}

\begin{document}

\title{Topological Atomic Spinwave Lattices by Dissipative Couplings}

\author{Dongdong Hao}%
\affiliation{Department of Physics, State Key Laboratory of Surface Physics and Key Laboratory of Micro and Nano Photonic Structures (Ministry of Education), Fudan University, Shanghai 200433, China}%
\author{Lin Wang}%
\affiliation{Department of Physics, State Key Laboratory of Surface Physics and Key Laboratory of Micro and Nano Photonic Structures (Ministry of Education), Fudan University, Shanghai 200433, China}%
\author{Xingda Lu}%
\affiliation{Department of Physics, State Key Laboratory of Surface Physics and Key Laboratory of Micro and Nano Photonic Structures (Ministry of Education), Fudan University, Shanghai 200433, China}%
\author{Xuzhen Cao}%
\affiliation{State Key Laboratory of Quantum Optics and Quantum Optics Devices, Institute of Laser Spectroscopy, Shanxi University, Taiyuan 030006, China}
\affiliation{Collaborative Innovation Center of Extreme Optics, Shanxi University, Taiyuan 030006, China}
\author{Suotang Jia}%
\affiliation{State Key Laboratory of Quantum Optics and Quantum Optics Devices, Institute of Laser Spectroscopy, Shanxi University, Taiyuan 030006, China}
\affiliation{Collaborative Innovation Center of Extreme Optics, Shanxi University, Taiyuan 030006, China}
\author{Ying Hu}%
\email{huying@sxu.edu.cn}
\affiliation{State Key Laboratory of Quantum Optics and Quantum Optics Devices, Institute of Laser Spectroscopy, Shanxi University, Taiyuan 030006, China}
\affiliation{Collaborative Innovation Center of Extreme Optics, Shanxi University, Taiyuan 030006, China}
\author{Yanhong Xiao}%
\email{yxiao@fudan.edu.cn}
\affiliation{State Key Laboratory of Quantum Optics and Quantum Optics Devices, Institute of Laser Spectroscopy, Shanxi University, Taiyuan 030006, China}
\affiliation{Collaborative Innovation Center of Extreme Optics, Shanxi University, Taiyuan 030006, China}
\affiliation{Department of Physics, State Key Laboratory of Surface Physics and Key Laboratory of Micro and Nano Photonic Structures (Ministry of Education), Fudan University, Shanghai 200433, China}%

\begin{abstract}
Recent experimental advance in creating dissipative couplings provides a new route for engineering exotic lattice systems and exploring topological dissipation. Using the spatial lattice of atomic spinwaves in a vacuum vapor cell, where purely dissipative couplings arise from diffusion of atoms, we experimentally realize a dissipative version of the Su-Schrieffer-Heeger (SSH) model. We construct the dissipation spectrum of the topological or trivial lattices via electromagnetically-induced-transparency (EIT) spectroscopy. The topological dissipation spectrum is found to exhibit edge modes within a dissipative gap. We validate chiral symmetry of the dissipative SSH couplings, and also probe topological features of the generalized dissipative SSH model. This work paves the way for realizing non-Hermitian topological quantum optics via dissipative couplings.
 \end{abstract}

\maketitle

\textit{Introduction --} Topological phases of quantum matter host fascinating phenomena such as edge modes that are immune to imperfections~\cite{Wen2004,HasanKaneReview2010,QiReview2011,Bernevigbook2013}, with potential applications in quantum computation and other technologies ~\cite{Kitaev2003,NayakReview2008,Pachos2012}. The robust nature of these phenomena in wide classes of lattice systems is linked to the presence of energy gaps and topologically nontrivial energy bands in the bulk; this protects edge modes, at energies within the bulk energy gap, from symmetry-preserving local perturbations.

Recently, dissipative couplings have been realized in various settings such as atoms~\cite{antiPT2016}, heat transfer system~\cite{heat-transfer2019}, circuits~\cite{circuits2018}, optomechanical systems~\cite{opto-mecha2021}, waveguides~\cite{waveguides2021}, resonators~\cite{resonators2019} and laser arrays~\cite{Davidson2021} etc. These advances opened up novel possibilities for designing topological structures~\cite{Diehl2011,Bardyn2013,Clerk2015,Fang2017,Wang2018,Kunst2018,Nunnenkamp2020,JanReview2021}. Lattice systems with purely dissipative couplings exhibit distinct spectral features from the coherently coupled networks in a Hamiltonian context and may enable topological dissipation, i.e., topological properties are associated with the gapped damping bands (or bands of dissipation rates) in the bulk, and dissipative edge modes within the dissipative gap, decoupled from the bulk. These intriguing phenomena, however, remain largely unexplored experimentally, with only a recent implementation using synthetic dimensions of photonic resonator with time-multiplexed pulses~\cite{FanSH2021}.

Atomic vapor systems offer a unique platform for exploring topological dissipation. Such systems involve a non-Markovian reservoir, where rapid transport of atomic coherence via atomic diffusions~\cite{NirRMP} naturally leads to dissipative coupling between long-lived atomic spinwaves created by electromagnetically-induced-transparency (EIT)~\cite{EIT-review2012} in spatially separated optical channels. This can mediate quantum optical spatial correlations as recently observed~\cite{CWX2020, LXD2021}. Realizing topological dissipation therein may promise topology-enabled quantum correlations and non-Hermitian topological quantum optics, complementary to topological quantum optics based on conservative couplings~\cite{Hafazi,Sergiv,WDW}.

Here we experimentally realize the dissipative version of the paradigmatic Su-Schrieffer-Heeger (SSH) model, based on a spatial lattice of atomic spinwaves in a vacuum vapor cell. Utility of the vacuum cell (i.e., no wall-coatings~\cite{Misha}) allows us to realize the nearest-neighbor dissipative couplings, hard to achieve in the wall-coated cell in previous experiments~\cite{antiPT2016} due to the all-to-all couplings therein. We control the coupling rates via the spacing between optical beams, thus inducing topological or trivial dissipation. By constructing the dissipation spectra via EIT spectroscopy, we show the topological dissipation spectrum exhibits edge modes at zero dissipation rates (relative to the background loss) within a bulk dissipative gap. We also create a ring pattern simulating a dissipative SSH model with periodic boundaries, and spectroscopically validate its chiral symmetry. Finally, we observe the weakly dissipative edge modes featured in the dissipative version of the generalized SSH model~\cite{Chenshu2014}. Our experiments agree well with the theoretical analysis.

\begin{figure}
	\centering
	\includegraphics[width=0.9\columnwidth]{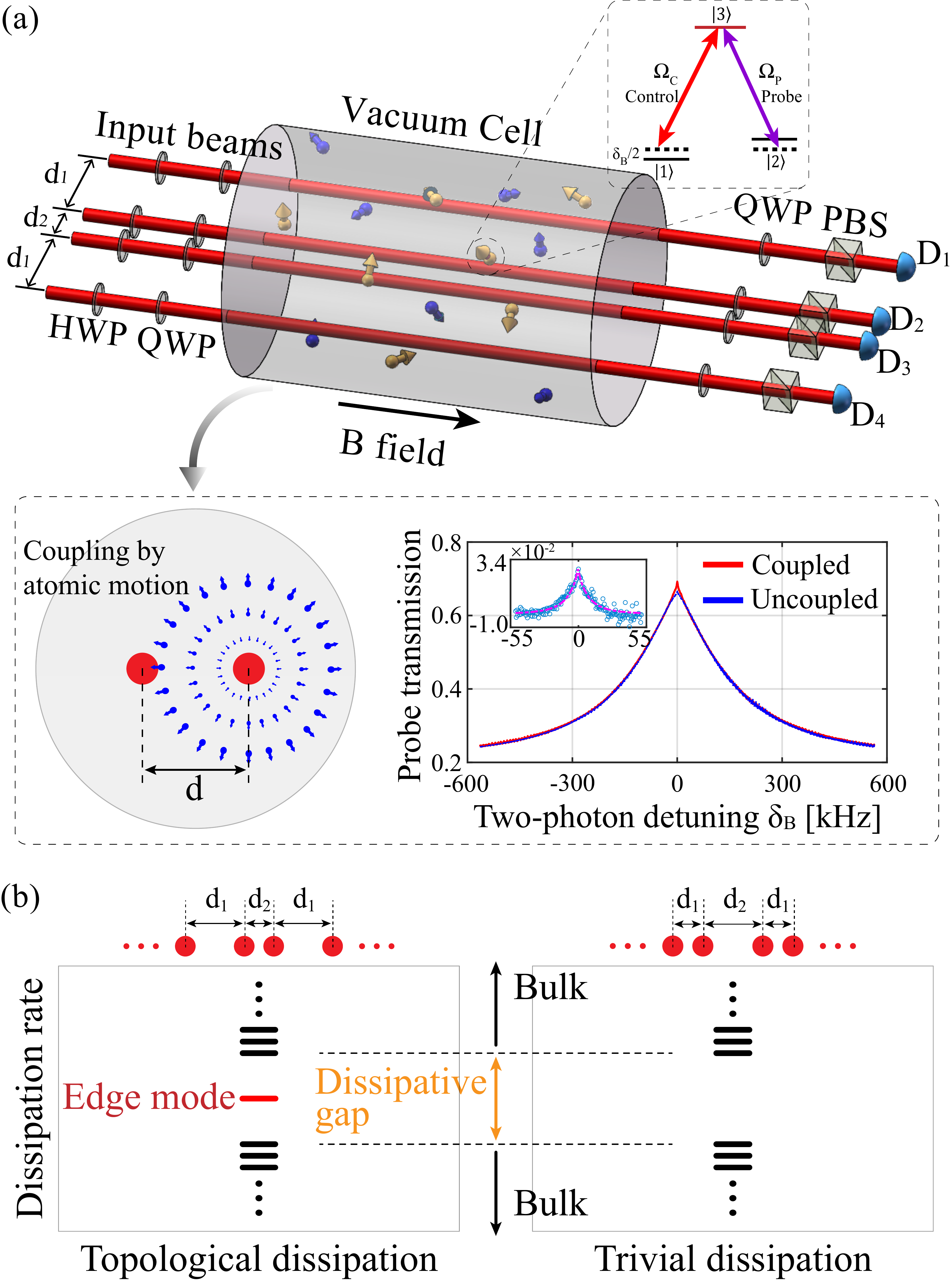}
	\caption{Schematics and principle of atomic vapor cell experiment simulating SSH model with dissipative couplings. (a) Setup. Top: Several optical channels with designed spacings in the vapor cell create ground state coherences (spinwaves) by EIT process in each channel. Spinwaves in neighboring channels couple to one another naturally through atomic motion, hence in a dissipative manner, with the coupling rate scaling as $1/d$. An array of spinwaves with alternating spacings $d_1$ and $d_2$ synthesizes the dissipative SSH model. The output probe intensities are measured to extract properties of the model. Bottom left: coupling mechanism. Bottom right: characterization of dissipative coupling rate using a two-channel setting; the dissipative coupling rate is measured from the difference of EIT spectra (inset) with and without inter-channel couplings~\cite{sup}. The dashed lines in the inset is the fitting (to guide the eye) of the experimental data shown as dots in the inset. (b) Schematic of the topological and trivial dissipation spectra expected from distinct patterns. Topological dissipation spectrum exhibits dissipative edge modes at isolated dissipation rates within a bulk dissipative gap, in contrast to the trivial spectrum. HWP: half wave plate, QWP: quarter wave plate, PBS: polarization beam splitter. }\label{Fig:setup}
\end{figure}

\textit{Dissipatively coupled SSH array by flying atoms --} Our experiments utilize an enriched $^{87}\text{Rb}$ vacuum vapor cell [Fig.~\ref{Fig:setup}(a)] with no buffer gas nor wall coating of cylindrical shape with a diameter of $2.5$ cm and length $5$ cm, and housed within a three-layer magnetic shield to screen out ambient magnetic fields. The cell temperature is set to $40^{\circ}$C to maintain a relatively small optical depth. The output of a diode laser, tuned to the Rb D1 transition $5S_{1/2}F=2\rightarrow5P_{1/2}F'=1$, passes through a polarization-maintaining optical fiber, and is then divided into several spatially separated beams, forming optical channels in the cell. Each channel is composed of a right-circularly polarized strong control field (Rabi frequency $\Omega_c$) and a left-circularly polarized weak probe ($\Omega_p$) forming a standard $\Lambda$-type EIT configuration with ground states $|1\rangle$,$|2\rangle$ and excited state $|3\rangle$, which creates a local atomic spinwave (i.e., ground-state coherence ${\rho}_{12}$). A uniform magnetic field is applied to induce Zeeman shifts to the energy levels for the adjustment of two-photon detuning $\delta_B$. Spinwaves in different optical channels are dissipatively coupled through atomic motion. The coupling rate is controlled through the channel separation $d$ with a $1/d$ scaling~\cite{sup}, while the laser beam diameter is set to $1.5$ mm for all the beams. \textcolor{black}{When channels are aligned in a straight line, the direct atom-flight path (for the ground state coherence) between the beyond-nearest-neighbor channels is largely ``blocked" via optical pumping of the channel(s) in between, and we effectively realize nearest-neighbor couplings.} Thus by patterning $2N$ channels with alternating spacings $d_1$ and $d_2$, we synthesize a dissipative form of the SSH model, with lattice site $j$ represented by the ground state coherence $\rho^{(j)}_{12}$ in channel $j$ ($j=1,...,2N$).

According to the standard density matrix formalism~\cite{sup}, our system can be described by the equation of motion $\partial_t\boldsymbol{\rho}_{12}=-i\left[(\delta_B-i\gamma)I+H\right]\boldsymbol{\rho}_{12}+\mathbf{P}_{in}$. Here, the vector $\boldsymbol{\rho}_{12}\equiv [\rho_{12}^{(1)},..., \rho_{12}^{(2N)}]^T$ denotes the ground state coherence distribution across the channels, $I$ is a unity matrix, $\gamma$ is the dephasing rate dominated by the transit broadening common to all channels, and the vector $\mathbf{P}_{in}\equiv [P_{in}^{(1)},...,P_{in}^{(2N)}]^T$ denotes the pumping sources of the coherence by the input light fields, where $\textrm{P}_{in}^{(j)}=-{\Omega^{(j)}_c}^{*}\Omega^{(j)}_p/\gamma_{23}$ with $\gamma_{23}$ the optical coherence decay rate. The non-Hermitian SSH Hamiltonian $H$ reads
\begin{eqnarray}
\!\!\!\!\!\!&&{H}= ive^{i\delta_\textrm{B} d_1/\nu}\sum_{m}\left(|a,m\rangle \langle b,m|+|b,m\rangle \langle a,m|\right)\nonumber\\
\!\!\!\!\!\!&+&iwe^{i\delta_\textrm{B} d_2/\nu}\sum_m \left(|b,m\rangle \langle a,m+1|+|a,m+1\rangle \langle b,m|\right),  \label{eq:H}
\end{eqnarray}
where $|a,m\rangle$ ($|b,m\rangle$) denotes $\rho^{(j)}_{12}$ in odd $j=2m-1$ (even $j=2m$) numbered channels in unit cell $m=1,...,N$. The intra- and inter-cell dissipative coupling rates $v$ and $w$ satisfy $v/w\propto d_2/d_1$. A relative phase $\delta_B(d_2-d_1)/\nu$ between the intra- and inter-cell couplings accumulates during the atomic flow (at velocity $\nu$) between neighboring beams and is the same in either directions.

The non-Hermitian Hamiltonian $H$ reduces to $H_0$ when $\delta_B=0$, realizing a purely dissipative version of the paradigmatic SSH model. It has chiral symmetry and inversion symmetry, and thus exhibits topological dissipation spectrum for $v<w$ [Fig.~\ref{Fig:setup}(b)], which features edge modes at zero dissipation rates in a bulk dissipative gap, absent for trivial dissipation where $v>w$.

\textit{EIT spectroscopy of topological dissipation spectrum -} We first implement a minimal version of the topological dissipative SSH model with $N=2$ unit cells in a geometry with open ends, using a chain of four laser beams with spacings $d_1=6$ mm and $d_2=3$ mm  [c.f. Fig.~\ref{Fig:setup}(a)]. The dissipative coupling rates are measured as $v\approx2\pi\times5$ kHz and $w\approx2\pi\times11$ kHz~\cite{sup}, with $v/w\approx d_2/d_1=1/2$ as expected. The background dephasing rate is measured as $\gamma\approx2\pi\times143$ kHz, which is barely affected by the inter-channel couplings because $ v,w\ll \gamma$.

We probe the non-Hermitian Hamiltonian $H$ via measuring the probe field's transmissions by sweeping $\delta_B$. Both the frequency and power of the laser are stabilized, and the laser polarization is carefully controlled. For an input $\mathbf{P}_{in}$, the output probe intensities are determined by the real part of the ground state coherences
\begin{eqnarray}
\boldsymbol{\rho}_{12}=-i((\delta_B-i\gamma)I+{H})^{-1}\mathbf{P}_{in}. \label{eq:EIT}
\end{eqnarray}
Thus information of $H$ is encoded in the \textit{difference} of $\boldsymbol{\rho}_{12}$ from $\boldsymbol{\rho}_{12}^0=-i(\delta_B-i\gamma)^{-1}{\bf P}_{in}$ in the uncoupled case (i.e., only the probe in the detected channel is on while all control fields are kept on). As the couplings are small and $\delta_B$-dependent, this difference is only significant in a narrow spectral window $|\delta_B|\lesssim v,w $ around the EIT center, and therefore, reflects essentially the purely dissipative case described by $H_0$ (i.e., when $\delta_B=0$).

\begin{figure}
\includegraphics[width=1\columnwidth]{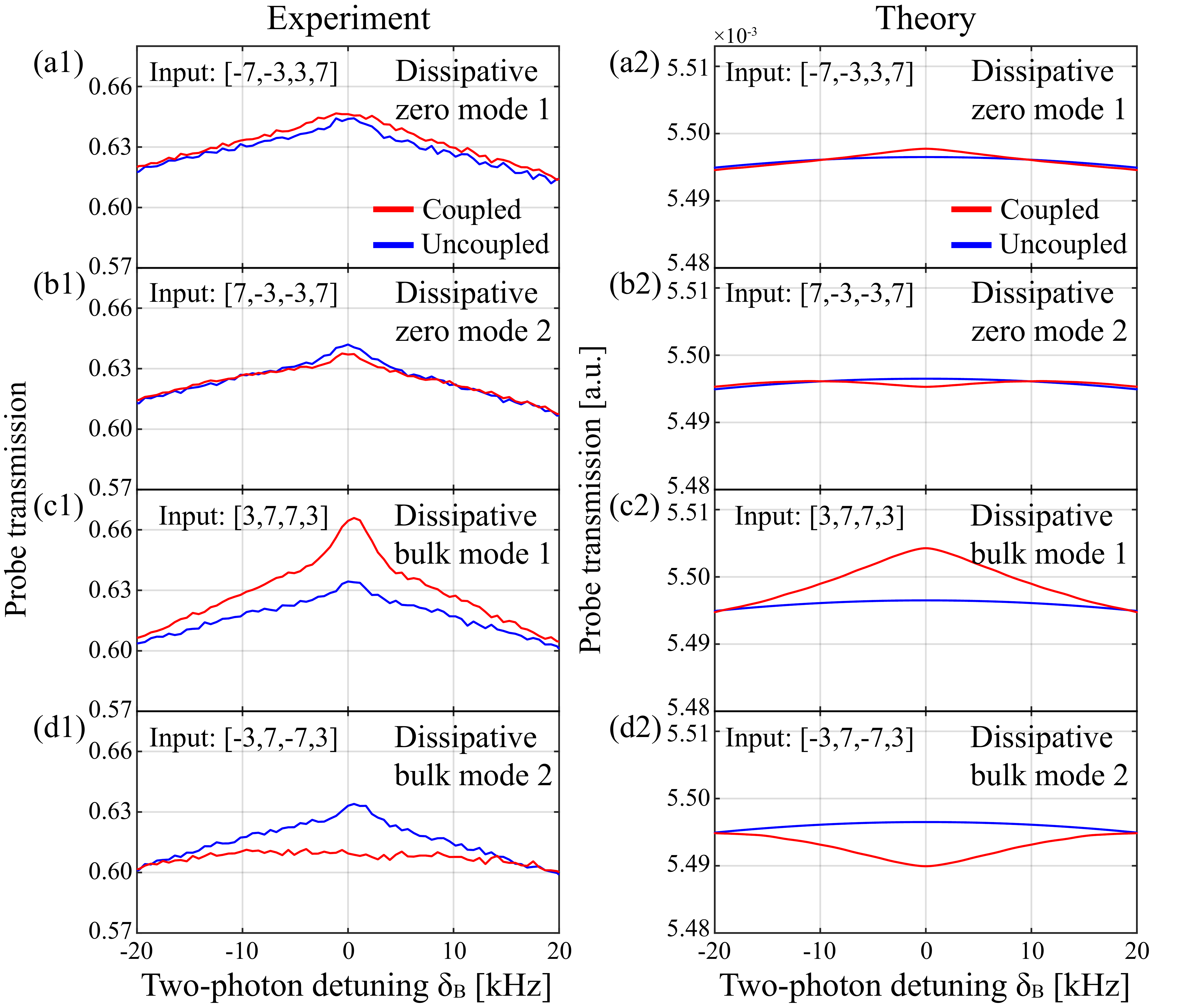}
\caption{\label{Fig:edge} Detections of the dissipation rates of edge and bulk modes in a topological dissipative SSH chain through eigen-EIT spectroscopy. The chain constitutes four channels in Fig.~\ref{Fig:setup}(a) with spacings $d_1=6$ mm and $d_2=3$ mm,  leading to dissipative coupling rates $v\approx2\pi\times5$kHz and $w\approx2\pi\times11$kHz. Only the center parts of the spectra manifesting influences from the dissipative SSH couplings are shown. (a1)-(d1) Measured eigen-EIT spectra via the probe transmission (normalized to far off resonant 100$\%$ transmission), coupled (all channel probes on) and uncoupled (only probe in the detected channel on), for the four input states $[-7, -3, 3, 7]^T$, $[7, -3, -3, 7]^T$, $[3, 7, 7, 3]^T$, $[-3, 7, -7, 3]^T$, respectively. The former (latter) two inputs approximate the two edge (bulk) states of the dissipative SSH model with dissipative coupling rates $v/w=1/2$. The difference between the coupled and uncoupled peak intensities provides the eigen-dissipation rates according to Eq.~(\ref{eq:eigenvalue}). Theoretical calculations of the optical coherences are shown in (a2)-(d2).}
\end{figure}
We detect the dissipation spectrum of $H_0$ via eigen-EIT spectroscopy. Let us label the eigen-dissipation rates of $H_0$ by $\gamma_\sigma$ ($\sigma=1,...,4$), which are defined by $H_0\psi_\sigma=i\gamma_\sigma\psi_\sigma$, where $\psi_\sigma$ denotes the corresponding eigenstates. To measure $\gamma_\sigma$, we harness the flexible control over the input light to design an eigenstate-form input $\mathbf{P}_{in}\propto \psi_\sigma$, which results in spinwaves and hence eigen-EIT supermode according to $\boldsymbol{\rho}_{12,\sigma}\propto\psi_\sigma/(-i\gamma+i\gamma_\sigma)$. Here, an eigenstate-form input is directly mapped to the spatial distribution of the output probe-laser power in each channel. After some algebra, we obtain the relation between the eigen-dissipation rates and the coupled and uncoupled $\rho_{12}$ in all the channels ($j=1,...4$):
\begin{equation}
\frac{{\rho}_{12,\sigma}^{(j)}-{\rho}_{12,\sigma}^{0,(j)}}{{\rho}_{12,\sigma}^{0,(j)}}\Big|_{\delta_B=0}=\frac{1}{1-\gamma_\sigma/\gamma}-1,\label{eq:eigenvalue}
\end{equation}
where $\boldsymbol{\rho}_{12,\sigma}^0\propto \psi_\sigma/(-i\gamma)$. Equation~(\ref{eq:eigenvalue}) is our central principle to accurately extract the dissipation spectrum.

\begin{figure}
	\centering
	\includegraphics[width=1\columnwidth]{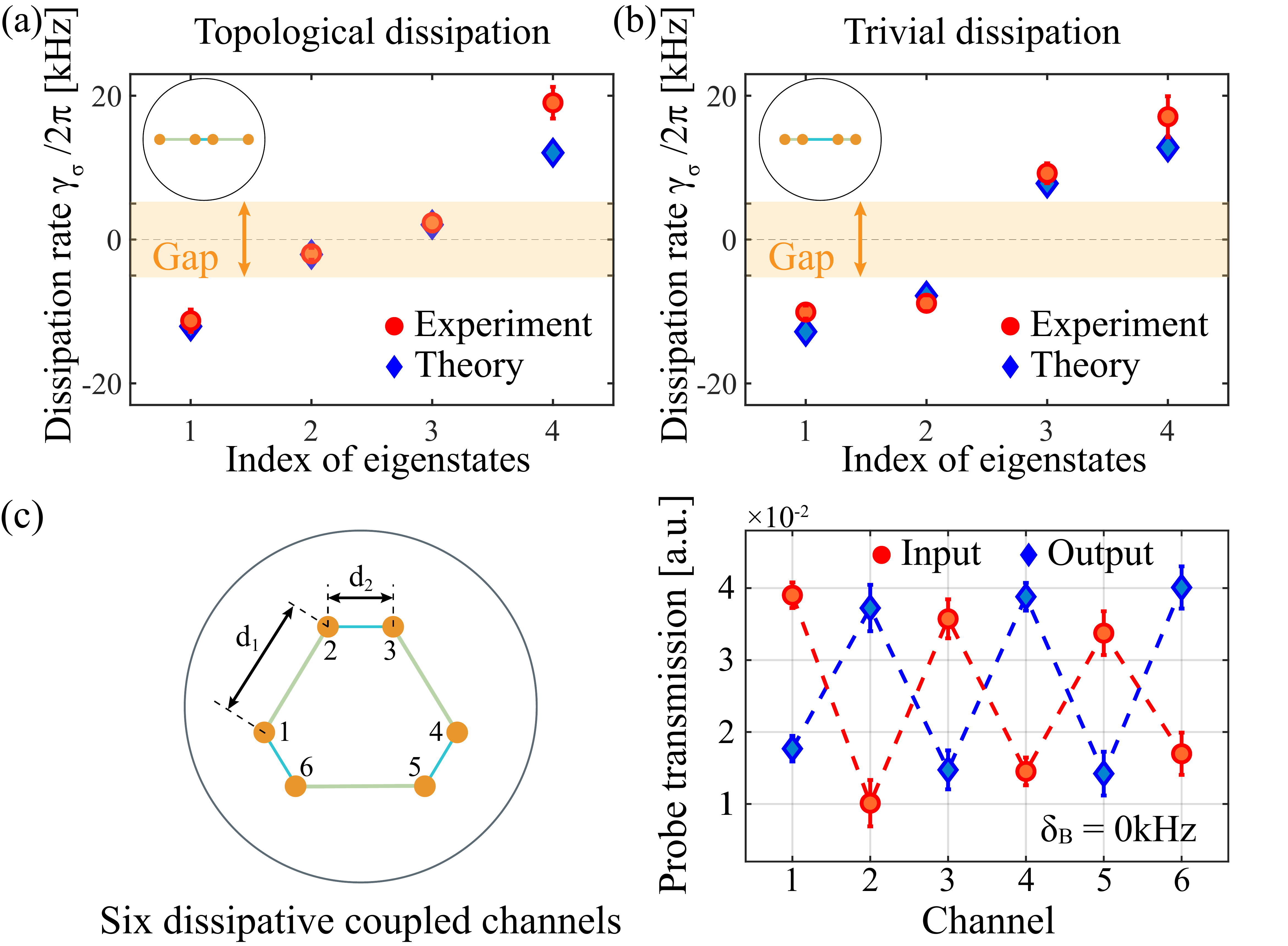}
	\caption{\label{Fig:spectrum} Observing features of topological dissipation. (a)-(b) Dissipation spectra for a dissipative SSH chain in (a) topological and (b) trivial regimes. The dissipation rates $\gamma_\sigma$, $\sigma=1,...,4$, are measured via eigen-EIT spectroscopy. The red dots denote the experimental data. The blue dots show predicted values from our theoretical model with (a) $v/w=1/2$ and (b) $v/w=2$. The orange region shows the expected dissipative gap.  The insets illustrate the cross-sections of (a) topological and (b) trivial patterns of four coupled optical channels in the vapor cell. (c) Chiral symmetry. Left panel illustrates the cross-section of $6$ channels forming a ring with spacings $d_1=6$ mm and $d_2=3$ mm. Right panel: The input probes (red) across the channels simulate $[1, 0, 1, 0, 1, 0]^T$. The ``output" probe laser intensities (blue) are the measured probe output differences between the coupled and uncoupled cases at $\delta_B=0$. In (a)-(c), the error bar is the standard deviation from about ten repetitive experiments.}
\end{figure}

Experimentally, we probe the four eigen-EIT supermodes in the lattice with $d_1/d_2\approx 2$ using four input vector states $\mathbf{P}_{in}$ denoted as $[-7, -3, 3, 7]^T$, $[7, -3, -3, 7]^T$, $[3, 7, 7, 3]^T$, $[-3, 7, -7, 3]^T$, respectively. These inputs resemble the theoretically-predicted eigenstates of $H_0$ with $v/w=1/2$: the former (latter) two mimic the edge (bulk) states. Here, numbers $``\pm7"$ and $``\pm3"$ represent the relative (approximate) values of ${\Omega}^*_c\Omega_p$ in each channel, $``7" (``3" )$ corresponds to a probe power of 9.5 $\mu$W (1.7 $\mu$W) while all control powers are fixed at 95.5 $\mu$W, and the $\pm$ sign is determined by the control and probe's relative phase, which are all judiciously set by tuning the wave plates in the light streams~\cite{sup}. \textcolor{black}{Since any channel yields the same eigenvalue according to Equation~(\ref{eq:eigenvalue}), we choose to measure the transmission spectra of the two channels with relatively higher probe power, thus higher signal to noise ratio, to extract an averaged eigenvalue.} Figures~\ref{Fig:edge}(a1)-(d1) show the four measured eigen-EIT spectra. We observe that the eigen-EIT spectra corresponding to the edge-state inputs nearly overlap with the uncoupled EIT [Figs.~\ref{Fig:edge}(a1)-(b1)], signaling the ``zero" eigen-dissipation rates (i.e., $\gamma_\sigma\approx 0$). In contrast, the bulk EIT supermodes in Figs.~\ref{Fig:edge}(c1)-(d1) change significantly in both the peak intensity and the linewidth compared to the uncoupled case; an increase (decrease) in the peak intensity indicates dissipation rates $\gamma_\sigma>0$ or $\gamma_\sigma<0$ following from Eq.~(\ref{eq:eigenvalue}). In Figs.~\ref{Fig:edge}(a2)-(d2), we present theoretical calculations~\cite{sup} using Eq.~(\ref{eq:H}) with measured $v$ and $w$ values. As shown, the relative trends in the coupled and uncoupled EIT spectra shown by the calculation are consistent with that in the experiment data, with the remaining disagreement in that experimental uncoupled EIT spectra are pointier than the theoretical ones, because the theory model gives an idealized Lorentzian lineshape which disregards experimental complications~\cite{sup}.

Finally we obtain the dissipation rates based on Eq.~(\ref{eq:eigenvalue}), using the measured peak intensities in the coupled and uncoupled cases in Figs.~\ref{Fig:edge}(a1)-(d1). The constructed dissipation spectrum for $v/w\approx 1/2$ is shown in Fig.~\ref{Fig:spectrum}(a). The key feature is the existence of two nearly zero dissipation rates deep within the expected dissipative gap of size $2|v-w|=2\pi\times12$ kHz; note the small degeneracy splitting is a natural consequence of the small system size here. By contrast, the other two dissipation rates are in the spectral bulk outside the gap.

To compare the spectra in topologically distinct phases, we swap the channel spacing to realize a configuration with $d_1=3$ mm and $d_2=6$ mm, corresponding to $v/w\approx 2$. Using eigen-EIT spectroscopy, we construct the dissipation spectrum [Fig.~\ref{Fig:spectrum}(b)]. All the dissipation rates are now outside the gap, in contrast to the topological spectrum [Fig.~\ref{Fig:spectrum}(a)]. For both, the dissipation rates distribute nearly symmetrically around zero. The experiment agrees with the calculations from diagonalizing $H_0$ with $v/w=1/2$ and $v/w=2$, respectively. \textcolor{black}{The discrepancy between the experiment and theory, especially for the largest eigenvalue, is due to the residual returned atomic coherence after wall collisions and other experiment imperfections~\cite{sup}.}

\textit{Probe chiral symmetry in a ring --} We next probe potential chiral symmetry of the dissipative SSH model~(\ref{eq:H}). To this end, we construct a ring configuration with $6$ laser beams [Fig.~\ref{Fig:spectrum}(c)] to implement the model with $N=3$ unit cells under periodic boundary condition. The chiral symmetric operator $S$ of an SSH ring is expressed as $S=I\otimes\sigma_z$, where $\sigma_z$ is the Pauli matrix. For the ring here, $I$ is a $3\times 3$ unity matrix. To experimentally probe chiral symmetry, we exploit the fact that the $S$ has two eigenvectors $S\phi_{\pm}=\pm \phi_{\pm}$, with $\phi_{+}=[1, 0, 1, 0, 1, 0]^T$ and $\phi_{-}=[0, 1, 0, 1, 0, 1]^T$ (states written for real space, unnormalized) corresponding to different chirality ($\pm$); if $H$ is chiral symmetric, i.e., $SHS^\dag=-H$, its action on, say $\phi_+$, yields the eigenstate with opposite chirality, $H\phi_{+}\propto \phi_{-}$. In this spirit, we prepare an input $\mathbf{P}_{in}\propto[1, 0, 1, 0, 1, 0]^T$ and measure the probe transmission change (due to couplings) in the $6$ channels as the output. Signature of chiral symmetry is observed [Fig.~\ref{Fig:spectrum}(c)]: the input in odd-numbered channels leads to an output dominantly in even-numbered channels. \textcolor{black}{We note that, although the presence of beyond-nearest-neighbor couplings breaks chiral symmetry in the strict sense, given that these coupling rates are smaller than $v$, $w$ ($\ll \gamma$) in our experiment, we are still able to observe residue signature of chiral symmetry in the transmission.}

\begin{figure}
	\centering
	\includegraphics[width=0.97\columnwidth]{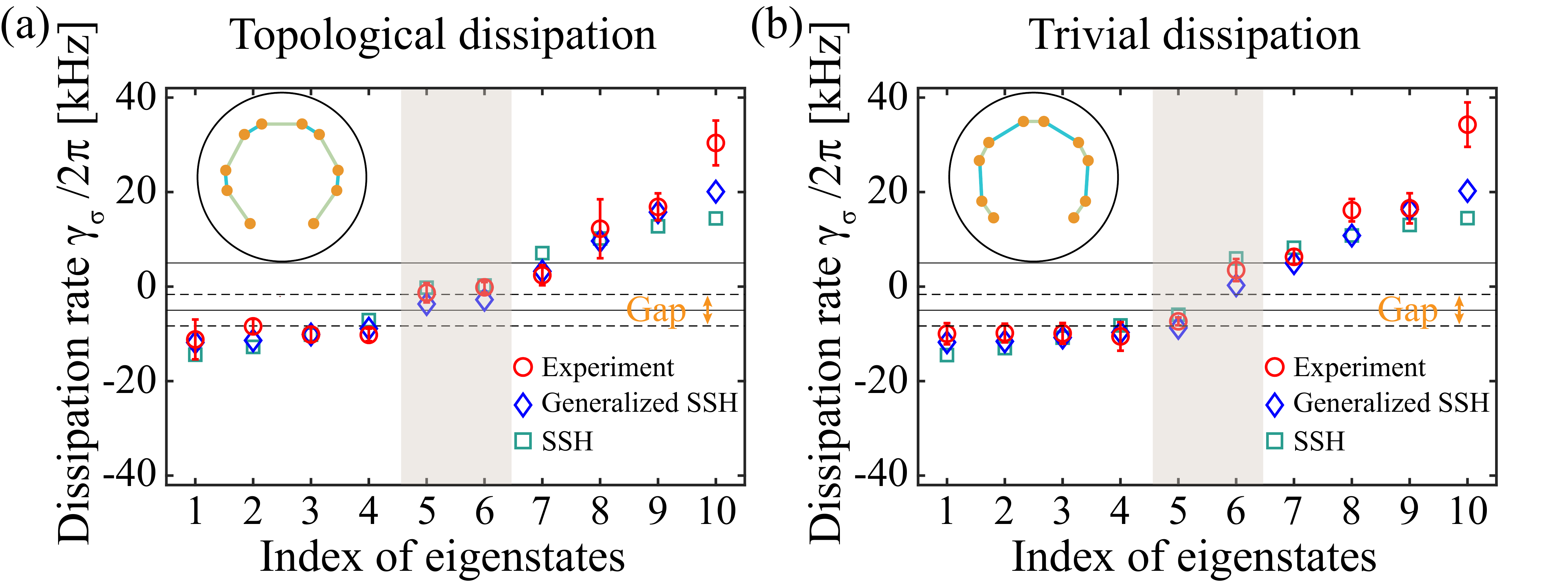}
	\caption{\color{black}\label{Fig:extend} Implementing the generalized dissipative SSH model with $10$ channels in (a) topological and (b) trivial regimes. Cross-sections of the open-end array are shown in the insets with (a) $d_1=6$ mm and $d_2=3$ mm, and (b) $d_1=3$ mm and $d_2=6$ mm. In the main panels, the red dots denote the experimental data. The blue diamonds denote the theoretical eigenvalues of the generalized dissipative SSH Hamiltonian (\ref{eq:HH}), and regions between the dashed lines indicate the gap. For comparison, the green squares denote the theoretical simulations from the dissipative SSH Hamiltonian $H$, and the gap is drawn by the solid lines. The vertical shaded area highlights the modes of particular interest (\textit{see text}). The error bar is the standard deviation from about ten repetitive experiments. The eigenstates (unnormalized) of the two edge modes are $[-0.56, 0.25, 0.24, -0.16, -0.21, 0.21, 0.16, -0.24, -0.25, 0.56]$, $[0.62, -0.17, -0.26, -0.02, 0.14, 0.14, -0.02, -0.26, -0.17,0.62]$ in terms of ground state coherence, corresponding to input probe powers of $[8.0, 1.6, 1.5, 0.7, 1.2, 1.2, 0.7, 1.5, 1.6, 8.0]$$\mu$W, $[10, 0.8, 1.7, 0.01, 0.5, 0.5, 0.01, 1.7, 0.8, 10]$$\mu$W respectively. The control beam powers in all channels are the same, $\sim95$ $\mu$W. Experiment results for the input states corresponding to the largest eigenvalues are affected the most by the couplings beyond the NNN, leading to greater discrepancy between the experiment and theory, because other input states have a more densely-interwoven distribution of positive and negative spin coherences which tends to average out the influences from sites further away. For instance, the 1st, 2nd, 8th, 9th and 10th eigenstates (unnormalized) in the topological regimes are respectively: $[-0.08, 0.26, -0.11, -0.49, 0.41, 0.41, -0.49, -0.11, 0.26, -0.08]$, $[0.07, -0.41, 0.38, 0.18, -0.39, 0.39, -0.18, -0.38, 0.41, -0.07]$, $[-0.31, -0.40, -0.29, 0.20, 0.35, 0.35, 0.20, -0.29, -0.40, -0.31]$, $[0.22, 0.40, 0.43, 0.29, 0.16, -0.16, -0.29, -0.43, -0.40, -0.22]$,
$[0.11, 0.25, 0.31, 0.39, 0.42, 0.42, 0.39, 0.31, 0.25, 0.11]$.}
\end{figure}

\color{black}\textit{Generalized dissipative SSH chain with $10$ channels --} In a ring-like pattern, direct atomic flight between the next-nearest channels may lead to nontrivial NNN coupling and modify the topological properties. To exam its effect, we wire up $10$ channels as an open-end chain with $d_1=6$ mm and $d_2=3$ mm [Fig.~\ref{Fig:extend} (a)], i.e., $v/w\approx 1/2$. The effective Hamiltonian including NNN couplings is
\begin{equation}
H'=H+H_\text{NNN}, \label{eq:HH}
\end{equation}
where $H_\text{NNN}=it \sum_{m}\big(|a,m\rangle \langle a,m+1|+|b,m\rangle \langle b,m+1|+\text{H.c}\big)$ captures dissipative NNN couplings with rate $t$ for $\delta_B\approx 0$. $H'$ represents the dissipative version of the so-called generalized SSH model~\cite{Chenshu2014}. Although $H'$ violates chiral symmetry~\cite{footnote4}, it retains inversion symmetry. Thus when $v/w\approx 1/2$, $H'$ is topological for $t/w<1/2$, which hosts two degenerate edge modes with dissipation rates $\gamma_\sigma<0$~\cite{sup}. Our realized array has $t/w\approx 1/3$ [Fig.~\ref{Fig:extend} (a)] and remains in the topological regime, but with a reduced gap $2t\approx 2\pi\times 6.67$ kHz. Note, $H'$ and $H$ share the same eigenstates in the bulk, but the spectrum of $H'$ is shifted by $\Delta\gamma_k=2t\cos k$ with respect to $H$ in the momentum space~\cite{footnote4}.

Experimentally, we measure the dissipation spectrum of $H'$ via the eigen-EIT spectroscopy, where the input states are engineered as the eigenstates of $H'$ with $v:w:t\approx \frac{1}{2}:1:\frac{1}{3}$. The experimental data are shown in Fig.~\ref{Fig:extend} (a). Furthermore, we swap the channel spacing to realize $H'$ with $v:w:t=1:\frac{1}{2}:\frac{1}{3}$ in the nontopological regime, and extract its spectrum as shown in Fig.~\ref{Fig:extend} (b). Each eignevalue is the averaged value of that measured from the transmission spectrum of the four channels with relatively higher probe powers. The experimental data are in good agreement with the calculated eigenvalues of $H'$. Although $H'$ with $t/w\approx 1/3$ is close to the phase boundary, comparison of Figs.~\ref{Fig:extend}(a) an (b)(especially the shaded regions) still allows to distinguish the nearly degenerate, surviving edge modes [see Fig.~\ref{Fig:extend}(a)] in the topological regime. Moreover, to extract the spectral shift due to NNN couplings, we compare the experiment with the theoretical simulation of the eigen-EIT transmission using Eq.~(\ref{eq:EIT}) with $H$ instead of $H'$. We observed the expected shift, e.g., both the maximum and minimum dissipation rates of $H'$, associated with $k=0$, shift upward. The remaining discrepancy between the experiment and theory is mainly attributed to the couplings beyond the NNN in the open-ring structure (see caption of Fig.~\ref{Fig:extend}).

\color{black}
\textit{Conclusions and outlook --} We have realized a lattice of atomic spinwaves with dissipative SSH couplings in a vapor cell and spectroscopically demonstrated key features of topological dissipation. Though the coupling strength remains small compared to the background loss, we expect to reach stronger coupling regime by engineering the geometry of a wall-coated cell, the laser beam profiles and their arrangements to prevent all-to-all coupling but still retain coherence protection by the coating. Other means to control the coupling could be incorporated, e.g., diffractive optical coupling~\cite{Davidson2021} and reservoir engineering~\cite{LXD2021}. Combining with the controllability over atomic spins by multi-level and nonlinear atom-light interactions in each channel~\cite{SunJ2019}, our platform holds unique promise for exploring non-Hermitian topology~\cite{JanReview2021,Wang2018,Kunst2018,Gong2018,Zhou2018,Kawabata2019,Xue2020,Yanbo2022} in quantum regimes, and designing novel quantum-correlated light sources for quantum information applications.


{\textit{Acknowledgment.}}  We are grateful to Wei Yi, Tao Shi, Shuai Chen, Heng Shen, Feng Mei, Changrui Yi and Ruiheng Jiao for discussions. This work is supported by the Natural Science Foundation of China (NSFC) under Grants No. 12161141018, No. 12027806, No. 61675047, and No. 11874038, and by the National Key Research and Development Program of China under Grants No. 2017YFA0304204.

\bibliographystyle{apsrev}

\begin{thebibliography}{99}
\bibitem{footnote4}{The Bloch Hamiltonian associated with $H'$ is $2it(\cos k) I+i(v+w\cos k)\sigma_x+iw(\sin k)\sigma_y$ with the unity matrix $I$, which shares the same eigenstates with the dissipative SSH model. Different from the latter, $H'$ with $v/w=1/2$ and $t/w=1/3$ exhibits an indirect gap: the minimum of the upper band and the maximum of the lower band are associated with different momenta. }
\end{thebibliography}

\end{document}


\title{Topological Atomic Spinwave Lattices by Dissipative Couplings}
\author{Dongdong Hao}%
\affiliation{Department of Physics, State Key Laboratory of Surface Physics and Key Laboratory of Micro and Nano Photonic Structures (Ministry of Education), Fudan University, Shanghai 200433, China}%
\author{Lin Wang}%
\affiliation{Department of Physics, State Key Laboratory of Surface Physics and Key Laboratory of Micro and Nano Photonic Structures (Ministry of Education), Fudan University, Shanghai 200433, China}%
\author{Xingda Lu}%
\affiliation{Department of Physics, State Key Laboratory of Surface Physics and Key Laboratory of Micro and Nano Photonic Structures (Ministry of Education), Fudan University, Shanghai 200433, China}%
\author{Xuzhen Cao}%
\affiliation{State Key Laboratory of Quantum Optics and Quantum Optics Devices, Institute of Opto-Electronics, Shanxi University, Taiyuan 030006, China}
\affiliation{Collaborative Innovation Center of Extreme Optics, Shanxi University, Taiyuan 030006, China}
\author{Suotang Jia}%
\affiliation{State Key Laboratory of Quantum Optics and Quantum Optics Devices, Institute of Laser Spectroscopy, Shanxi University, Taiyuan 030006, China}
\affiliation{Collaborative Innovation Center of Extreme Optics, Shanxi University, Taiyuan 030006, China}
\author{Ying Hu}%
\email{huying@sxu.edu.cn}
\affiliation{State Key Laboratory of Quantum Optics and Quantum Optics Devices, Institute of Laser Spectroscopy, Shanxi University, Taiyuan 030006, China}
\affiliation{Collaborative Innovation Center of Extreme Optics, Shanxi University, Taiyuan 030006, China}
\author{Yanhong Xiao}%
\email{yxiao@fudan.edu.cn}
\affiliation{State Key Laboratory of Quantum Optics and Quantum Optics Devices, Institute of Laser Spectroscopy, Shanxi University, Taiyuan 030006, China}
\affiliation{Collaborative Innovation Center of Extreme Optics, Shanxi University, Taiyuan 030006, China}
\affiliation{Department of Physics, State Key Laboratory of Surface Physics and Key Laboratory of Micro and Nano Photonic Structures (Ministry of Education), Fudan University, Shanghai 200433, China}%
\maketitle
\setcounter{equation}{0}
\setcounter{figure}{0}
\renewcommand{\theequation}{S\arabic{equation}}
\renewcommand{\thefigure}{S\arabic{figure}}
\renewcommand{\thetable}{S\Roman{table}}
\renewcommand{\bibnumfmt}[1]{[S#1]}
\renewcommand{\citenumfont}[1]{S#1}

\section{Theoretical model}

\begin{figure}
	\centering
	\includegraphics[width=0.7\textwidth]{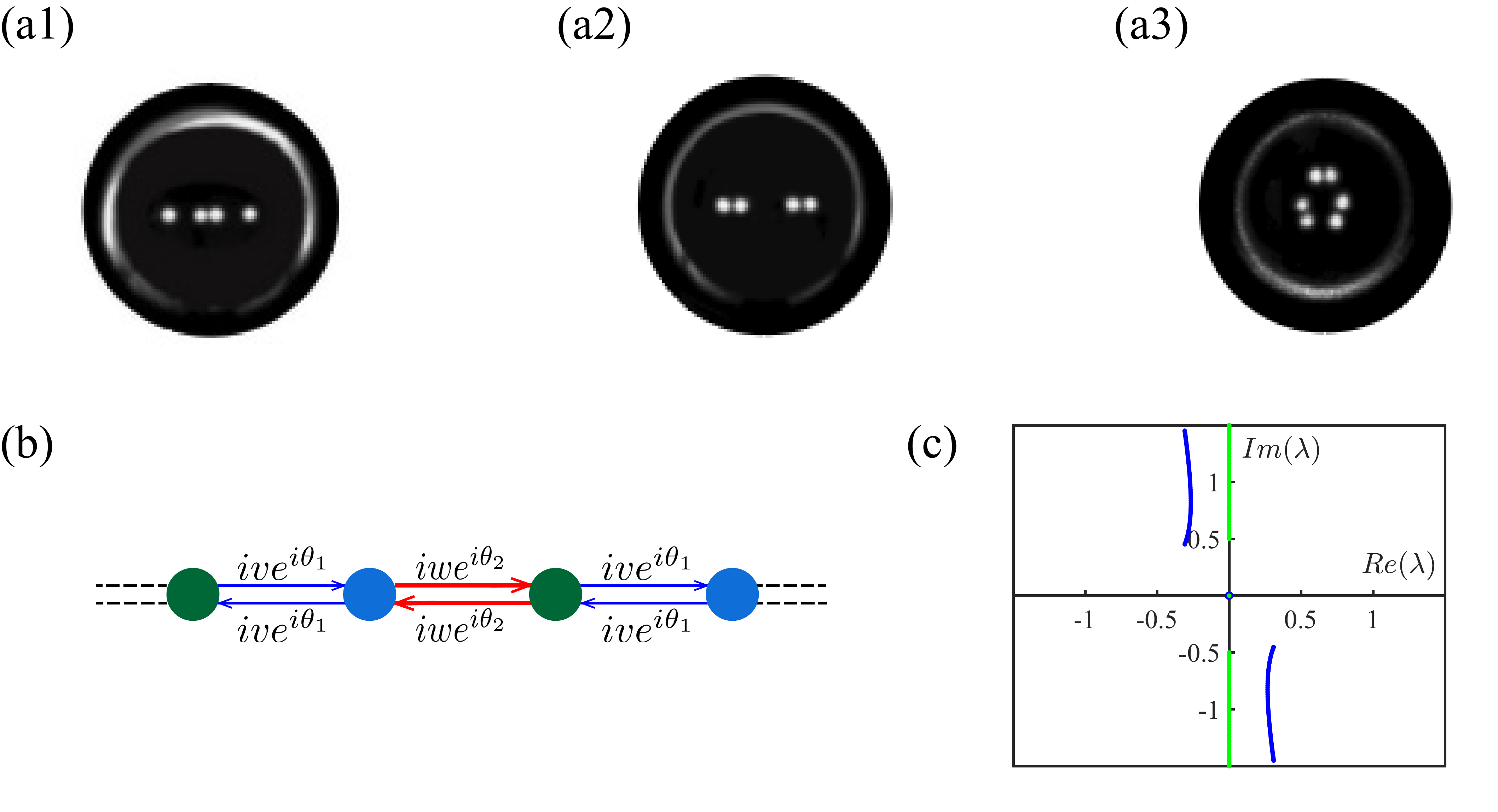}
	\caption{Schematics and principle of atomic vapor cell experiment simulating SSH model with dissipative couplings. (a) Experimentally implemented spin chain configurations shown by the optical channels, where each channel is a dot in the picture by a camera. (a1) and (a2) are four-channel open-end chains, and (a3) six-channel ring. (b) Schematics of the non-Hermitian SSH model with dissipative couplings. (c) Calculated eigenvalue spectra with $\theta=0$ (green) and $\theta=0.1\pi$ (blue), taking $v/w=1/2$ in the topological phase. The eigenvalues $\lambda$ are shown in the complex plane. In both cases, there exist zero modes, with both zero dissipation rate and zero energy.}
\label{Fig. S1}
\end{figure}

The spin wave chain is formed by an array of optical channels in the vapor cell, where each channel contains spatially overlapping control and probe fields that write ground atomic coherence through electromagnetically induced transparency (EIT). The laser fields propagate along the long axis of the cylindrical vapor cell, and various types of spin wave chains can be formed simply by arranging the position of the laser beams. Shown in Fig.~\ref{Fig. S1} are the cell-cross-section pictures (taken by a camera) of three types of chains used in our experiment.

We have developed a simplified theoretical model to describe the coupling and dynamics of the atomic coherence in the optical channels, which is an extension of our previous model for anti-PT symmetry with flying atoms in a wall-coated vapor cell ~\cite{antiPT2016}. The main difference is that, here, in the vacuum cell without wall coating, the coupling rate between any two optical channels depends on the channel spacing, while in a (cylindrical) coated cell the coupling rates between any two channels of the same laser diameters are the same.



Using similar procedures as described in ref~\cite{antiPT2016}, we derive the effective coupling Hamiltonian for the four- or six-channel case in a vacuum vapor cell by solving the master equation Eq. \eqref{S1}. For simplicity, we assume an optically thin medium and thus neglect the propagation effect.
\begin{equation}
\label{S1}
\tag{S1}
\begin{aligned}
\begin{split}
&\dot{\rho}^{(1)} + \Gamma^{(1)}_{rel}\rho^{(1)} = -i [H^{(1)}, \rho^{(1)}] + \Gamma^{(1)}_{exc}\rho^{(1)} + ve^{i\theta_{1}}\rho^{(2)}\\
&\dot{\rho}^{(2)} + \Gamma^{(2)}_{rel}\rho^{(2)} = -i [H^{(2)}, \rho^{(2)}] + \Gamma^{(2)}_{exc}\rho^{(2)} + ve^{i\theta_{1}}\rho^{(1)} + we^{i\theta_{2}}\rho^{(3)}\\
&\dot{\rho}^{(3)} + \Gamma^{(3)}_{rel}\rho^{(3)} = -i [H^{(3)}, \rho^{(3)}] + \Gamma^{(3)}_{exc}\rho^{(3)} + ve^{i\theta_{1}}\rho^{(4)} + we^{i\theta_{2}}\rho^{(2)}\\
&\dot{\rho}^{(4)} + \Gamma^{(4)}_{rel}\rho^{(4)} = -i [H^{(4)}, \rho^{(4)}] + \Gamma^{(4)}_{exc}\rho^{(4)} + ve^{i\theta_{1}}\rho^{(3)}\\
\end{split}
\end{aligned}
\end{equation}

\begin{figure}
	\centering
	\includegraphics[width=0.87\textwidth]{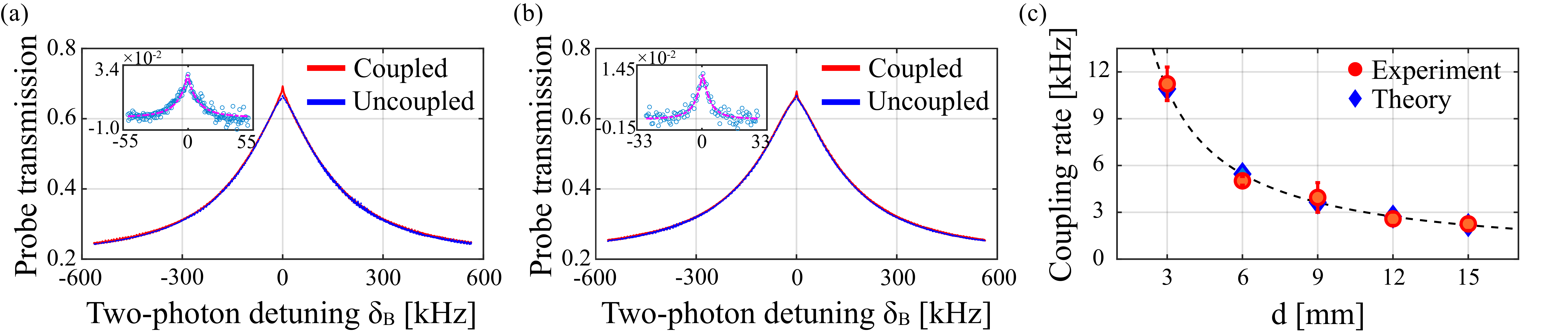}
	\caption{Characterization of dissipative coupling rate using a two-channel setting. (a) and (b) are EIT spectra with and without inter-channel couplings, for channel spacing of 3 mm and 6 mm respectively. The difference of the coupled (red) and uncoupled (blue) EIT is shown in the inset, whose linewidth is theoretically equal to the coupling rate as indicated by our numerical calculations. The full linewidth (at half maximum) of the EIT in the insets are about 11 kHz and 5 kHz, for $d=3$ mm and 6 mm respectively. The dashed lines in the insets are the fits (to guide the eye) of the experimental data shown as dots in the inset. Plotted in (c) is the coupling rates dependence on the channel spacing $d$, displaying an expected $\frac{1}{d}$ dependence. Here, the relative coupling rates are measured using a similar method as described in Eq.(3) of the main text, and then their absolute values are acquired by the difference EIT spectrum's linewidth.}
\label{Fig. S2}
\end{figure}Here, $\hbar = 1$ is assumed. $\Gamma^{(i)}_{rel}$ $(i=1,2,3,4)$ is the relaxation matrix representing the decays in the $i$th channel, and $\Gamma^{(i)}_{exc}$ stands for the excitation of the coherence and population. We set all laser beam diameters to be 1.5 mm and the distance between channels $d_{1} = 6$ mm, $d_{2} = 3$ mm, with $d_{1}$ the distance between channel 1 and 2 (also channel 3 and 4), $d_{2}$ the distance between channel 2 and 3 (see Fig.1 in the main text). While traveling from one channel to another, the ground state atomic coherence picks up a phase $\theta_{1}=\delta_B d_{1}/\nu$ or $\theta_{2}=\delta_B d_{2}/\nu$ with $\delta_B$ the two-photon detuning and $\nu$ the velocity of the atoms. The ground state coherence in four channels couple with one another through atomic motion at rates $v$ ($\propto \frac{1}{d_{1}}$) and $w$ ($\propto \frac{1}{d_{2}}$), and $|w|/|v|\approx2$ for the above settings of $d_{1}$ and $d_{2}$ (as verified in Fig.~\ref{Fig. S2}). After the rotating-wave approximation, the atom-light interaction for the four-channel case can be described by the following density-matrix equations:
\begin{equation}
	\label{S2}
	\tag{S2}
	\begin{aligned}
		\begin{split}
			&\dot{\rho}^{(1)}_{12} = -(\gamma_{12} + i\delta_B)\rho^{(1)}_{12} + i{\Omega^{(1)}_{c}}^{*}\rho^{(1)}_{32} - i\Omega^{(1)}_{p}\rho^{(1)}_{13} + ve^{i\theta_{1}}\rho^{(2)}_{12}\\
			&\dot{\rho}^{(1)}_{13} = -(\Gamma/2 + i\Delta^{(1)}_{1})\rho^{(1)}_{13} + i{\Omega^{(1)}_{c}}^{*}(\rho^{(1)}_{33}-\rho^{(1)}_{11}) - i{\Omega^{(1)}_{p}}^{*}\rho^{(1)}_{12} \\
			&\dot{\rho}^{(1)}_{32} = -(\Gamma/2 - i\Delta^{(1)}_{2})\rho^{(1)}_{32} - i\Omega^{(1)}_{p}(\rho^{(1)}_{33}-\rho^{(1)}_{22}) + i\Omega^{(1)}_{c}\rho^{(1)}_{12} \\ 	
			&\dot{\rho}^{(2)}_{12} = -(\gamma_{12} + i\delta_B)\rho^{(2)}_{12} + i{\Omega^{(2)}_{c}}^{*}\rho^{(2)}_{32} - i\Omega^{(2)}_{p}\rho^{(2)}_{13} + ve^{i\theta_{1}}\rho^{(1)}_{12} + we^{i\theta_{2}}\rho^{(3)}_{12}\\
			&\dot{\rho}^{(2)}_{13} = -(\Gamma/2 + i\Delta^{(2)}_{1})\rho^{(2)}_{13} + i{\Omega^{(2)}_{c}}^{*}(\rho^{(2)}_{33}-\rho^{(2)}_{11}) - i{\Omega^{(2)}_{p}}^{*}\rho^{(2)}_{12} \\
			&\dot{\rho}^{(2)}_{32} = -(\Gamma/2 - i\Delta^{(2)}_{2})\rho^{(2)}_{32} - i\Omega^{(2)}_{p}(\rho^{(2)}_{33}-\rho^{(2)}_{22}) + i\Omega^{(2)}_{c}\rho^{(2)}_{12} \\
			&\dot{\rho}^{(3)}_{12} = -(\gamma_{12} + i\delta_B)\rho^{(3)}_{12} + i{\Omega^{(3)}_{c}}^{*}\rho^{(3)}_{32} - i\Omega^{(3)}_{p}\rho^{(3)}_{13} + ve^{i\theta_{1}}\rho^{(4)}_{12} + we^{i\theta_{2}}\rho^{(2)}_{12}\\
			&\dot{\rho}^{(3)}_{13} = -(\Gamma/2 + i\Delta^{(3)}_{1})\rho^{(3)}_{13} + i{\Omega^{(3)}_{c}}^{*}(\rho^{(3)}_{33}-\rho^{(3)}_{11}) - i{\Omega^{(3)}_{p}}^{*}\rho^{(3)}_{12} \\
			&\dot{\rho}^{(3)}_{32} = -(\Gamma/2 - i\Delta^{(3)}_{2})\rho^{(3)}_{32} - i\Omega^{(3)}_{p}(\rho^{(3)}_{33}-\rho^{(3)}_{22}) + i\Omega^{(3)}_{c}\rho^{(3)}_{12} \\
			&\dot{\rho}^{(4)}_{12} = -(\gamma_{12} + i\delta_B)\rho^{(4)}_{12} + i{\Omega^{(4)}_{c}}^{*}\rho^{(4)}_{32} - i\Omega^{(4)}_{p}\rho^{(4)}_{13} + ve^{i\theta_{1}}\rho^{(3)}_{12}\\
			&\dot{\rho}^{(4)}_{13} = -(\Gamma/2 + i\Delta^{(4)}_{1})\rho^{(4)}_{13} + i{\Omega^{(4)}_{c}}^{*}(\rho^{(4)}_{33}-\rho^{(4)}_{11}) - i{\Omega^{(4)}_{p}}^{*}\rho^{(4)}_{12} \\
			&\dot{\rho}^{(4)}_{32} = -(\Gamma/2 - i\Delta^{(4)}_{2})\rho^{(4)}_{32} - i\Omega^{(4)}_{p}(\rho^{(4)}_{33}-\rho^{(4)}_{22}) + i\Omega^{(4)}_{c}\rho^{(4)}_{12} \\
		\end{split}
	\end{aligned}
\end{equation}
Here, $\gamma_{12}$ is the decay rate of the ground state coherence, $\Gamma$ is the decay rate of the excited state $\ket{3}$, $\Delta^{(i)}_{1}$ and $\Delta^{(i)}_{2}$ are one-photon detunings for the control and probe fields, and $\Omega^{(i)}_{c}$ and $\Omega^{(i)}_{p}$ are Rabi frequencies for the control and probe fields in the $i$th channel respectively.
Since the optical coherences $\rho^{(i)}_{13}$ and $\rho^{(i)}_{32}$ decay much faster than the ground-state coherences $\rho^{(i)}_{12}$, we can assume that they adiabatically follow the ground-state coherences. Therefore, by setting the time derivatives of optical coherences in Eq.\eqref{S2} to be zero, one can express the optical coherences in terms of the ground-state coherences. Also, in the experiment, we have $\Omega^{(i)}_{c} \gg \Omega^{(i)}_{p} $, and the control field itself is weak, which allow for the assumptions that the excited-state population $\rho_{33} = 0$, and one of the ground state population $\rho_{22} = 1$. We then arrive at the following coupled equations for the time evolution of the ground-state coherences of the four-channel case:
\begin{equation}
\label{S3}
\tag{S3}
\begin{aligned}
\begin{split}
	&\dot{\rho}^{(1)}_{12} = -(\gamma + i\delta_B)\rho^{(1)}_{12} + ve^{i\theta_{1}}\rho^{(2)}_{12} - \dfrac{{\Omega^{(1)}_{c}}^{*}\Omega^{(1)}_{p}}{\gamma_{23}}\\	
	&\dot{\rho}^{(2)}_{12} = -(\gamma + i\delta_B)\rho^{(2)}_{12} + ve^{i\theta_{1}}\rho^{(1)}_{12} + we^{i\theta_{2}}\rho^{(3)}_{12} - \dfrac{{\Omega^{(2)}_{c}}^{*}\Omega^{(2)}_{p}}{\gamma_{23}}\\
	&\dot{\rho}^{(3)}_{12} = -(\gamma + i\delta_B)\rho^{(3)}_{12} + ve^{i\theta_{1}}\rho^{(4)}_{12} + we^{i\theta_{2}}\rho^{(2)}_{12} - \dfrac{{\Omega^{(3)}_{c}}^{*}\Omega^{(3)}_{p}}{\gamma_{23}}\\
	&\dot{\rho}^{(4)}_{12} = -(\gamma + i\delta_B)\rho^{(4)}_{12} + ve^{i\theta_{1}}\rho^{(3)}_{12} - \dfrac{{\Omega^{(4)}_{c}}^{*}\Omega^{(4)}_{p}}{\gamma_{23}}\\
\end{split}
\end{aligned}
\end{equation}
Here, $\gamma = \gamma_{12} + \Gamma^{i}_{p}$ represents the total effective decay rate of the ground state coherence, with $\Gamma^{i}_{p} = |\Omega^{i}_{c}|^2 / \gamma_{23}$ ($i =1,2,3,4$) the optical pumping rate. From Eq.\eqref{S3}, one can deduce the effective Hamiltonian $\hat{H}_{eff}$ that depicts the coupling between the coherences in the four channels:

\begin{equation}
	\label{S4}
	\tag{S4}
	\begin{aligned}
		\begin{split}
			\hat{H_{eff}} &= \hat{H} + \hat{H}_{onsite} \\
			&=
			\begin{pmatrix}
				0 & ive^{i\theta_{1}} & 0 & 0\\
				ive^{i\theta_{1}} & 0 & iwe^{i\theta_{2}} & 0\\
				0 & iwe^{i\theta_{2}} & 0 & ive^{i\theta_{1}}\\
				0 & 0 & ive^{i\theta_{1}} & 0 \\
			\end{pmatrix}
			+(\delta_B-i\gamma)I
		\end{split}
	\end{aligned}
\end{equation}
Here, $\hat{H}$ characterizes a purely dissipative one-dimension SSH chain with the nearest neighbor coupling, $\hat{H}_{onsite}$ is the decay in local channels corresponding to the broad structure of the EIT spectrum and $I$ is the identity matrix. When the control and probe fields in all channels are on, the system is in a ``coupled" condition, and the change in the EIT spectra compared to the uncoupled case (with all control fields on but only one probe field on) is caused by $\hat{H}$. In general, the coupled EIT spectrum in each channel features a dual structure: a small but narrow peak (or dip, depending on the relative phase of the ground state coherences in different channels) at the central top of the uncoupled EIT spectra. 

We emphasize that the SSH Hamiltonian $\hat{H}$ has pure imaginary coupling terms only when the two-photon detuning $\delta_B$ is zero, otherwise the coupling terms are complex numbers, resulting in a more general non-Hermitian Hamiltonian. In this general scenario, chiral symmetry still exists, since $\hat{H}$ in momentum space below commutes with $\sigma_{z}$ for arbitrary $\delta_B$,
\begin{equation}
\label{S5}
\tag{S5}
	H(k)=
	\begin{pmatrix}
		0 & ive^{i\theta_{1}}+iwe^{i(\theta_{2}-k)} \\
		ive^{i\theta_{1}}+iwe^{i(\theta_{2}+k)} & 0
	\end{pmatrix}
\end{equation}
In the spatial domain, for the six-channel experiment with a coherence input of $[1,0,1,0,1,0]^T$, the coherence difference between the coupled and uncoupled cases are given by the following equation,
\begin{equation}
\label{S6}
\tag{S6}
\begin{aligned}
\begin{split}
\begin{pmatrix}
	0 & ve^{i\theta_{1}} & 0 & 0 & 0 & we^{i\theta_{2}}\\
	ve^{i\theta_{1}} & 0 & we^{i\theta_{2}} & 0 & 0 & 0\\
	0 & we^{i\theta_{2}} & 0 & ve^{i\theta_{1}} & 0 & 0\\
	0 & 0 & ve^{i\theta_{1}} & 0 & we^{i\theta_{2}} & 0\\
	0 & 0 & 0 & we^{i\theta_{2}} & 0 & ve^{i\theta_{1}}\\
	we^{i\theta_{2}} & 0 & 0 & 0 & ve^{i\theta_{1}} & 0\\
\end{pmatrix}
\begin{pmatrix}
	1\\
	0\\
	1\\
	0\\
	1\\
	0\\
\end{pmatrix}
= (ve^{i\theta_{1}} + we^{i\theta_{2}})
\begin{pmatrix}
	0\\
	1\\
	0\\
	1\\
	0\\
	1\\
\end{pmatrix}
\end{split}
\end{aligned}
\end{equation}
where the real part of the right hand side of Eq.\eqref{S6} determines the intensity change of the probe field upon coupling, and is consistent with the experiment observation as shown in Fig.4(c) in the main text.

\section{EIT spectroscopy for topology phase probing}

In our experiment, atoms in the vapor cell move fast ($\sim210$ m/s) and the atom-light interaction time is relatively short, giving rise to a transit decay ($\sim143$ kHz of linewidth broadening) of the local spin wave in all the optical channels, much larger than the coupling rate between the optical channels. Such loss makes it hard to observe the eigen-EIT-modes associated with the relatively weak SSH-coupling Hamiltonian. We propose to experimentally prepare and inject the eigen-EIT-modes predicted by the SSH Hamiltonian (with coupling rates from the experiment). From Eq.\eqref{S3}, the steady state ground state coherence (uncoupled case) is $\rho^{(i)}_{12}\propto{\Omega^{(i)}_{c}}^{*}\Omega^{(i)}_{p}$, which indicates that an ``input" state, i.e., a particular coherence distribution in all channels without the inter-channel coupling, can be prepared by adjusting the laser power of the probe fields (while maintaining the condition of stronger control and weaker probe) and the relative phase between the control and probe.

The experiment procedure of preparing the eigen-state input is as follows. First, the eigenstates and eigenvalues of the purely dissipative Hamiltonian $\hat{H}_0$ is calculated (by setting $\delta_B=0$ and $|w|/|v|=2$) as $[3,7,7,3]^T$, $[-3,7,-7,3]^T$, $[-7,-3,3,7]^T$ and $[7,-3,-3,7]^T$ (un-normalized), which corresponds to eigenvalues 0.97, $-0.97$, 0.17 and $-0.17$ respectively. Then, we construct a particular coherence distribution in the four channels by choosing the correct probe field powers while keeping the control powers unchanged. The polarization state of light can be described by a vector $\cos(\theta/2)\ket{R}+e^{i\phi}\sin(\theta/2)\ket{L}$ on the Poincare sphere, where $\theta$ is the polar angle and $\phi$ is the azimuthal angle, which determine the power ratio and relative phase between the right-handed circularly polarized control field $\ket{R}$ and the left-handed circularly polarized probe field $\ket{L}$. $\theta$ and $\phi$ are adjusted by half-wave and quarter-wave plates in the light streams before the vacuum cell.


The SSH coupling Hamiltonian determines the difference between the ground state coherence under the uncoupled and coupled conditions, and such difference can be measured through the probe transmission change on EIT resonance, since the ground state coherence is connected with the optical coherence by:
\begin{equation}
	\label{S7}
	\tag{S7}
	\rho^{(i)}_{32} = \dfrac{2i\Omega^{(i)}_{p}}{\Gamma} (1 + \dfrac{ \Omega^{(i)}_{c} }{\Omega^{(i)}_{p}} \rho^{(i)}_{12})\\
\end{equation}
and the probe transmission is given by the optical coherence through:
\begin{equation}
	\label{S8}
	\tag{S8}
	T^{(i)}_{p} = \exp(-\alpha L\dfrac{Im(\rho^{(i)}_{32})}{\Omega^{(i)}_{p}})\\
\end{equation}
Here, $L$ is the length of the vacuum cell, $\alpha = 2n\mu_{0}^2/\lambda\epsilon_{0}\hbar$, with $n$ the atomic number density, $\mu_{0}$ the dipole moment, $\lambda$ the wavelength of the input light, $\epsilon_{0}$ the permittivity of vacuum.
To be more precise, we take the logarithm of the normalized EIT spectra to obtain the (average) optical coherence, although the optical depth is relatively small and taking the logarithm has a small effect on the results.


\section{Extended experiment data}

\begin{figure}
	\centering
	\includegraphics[width=0.5\textwidth]{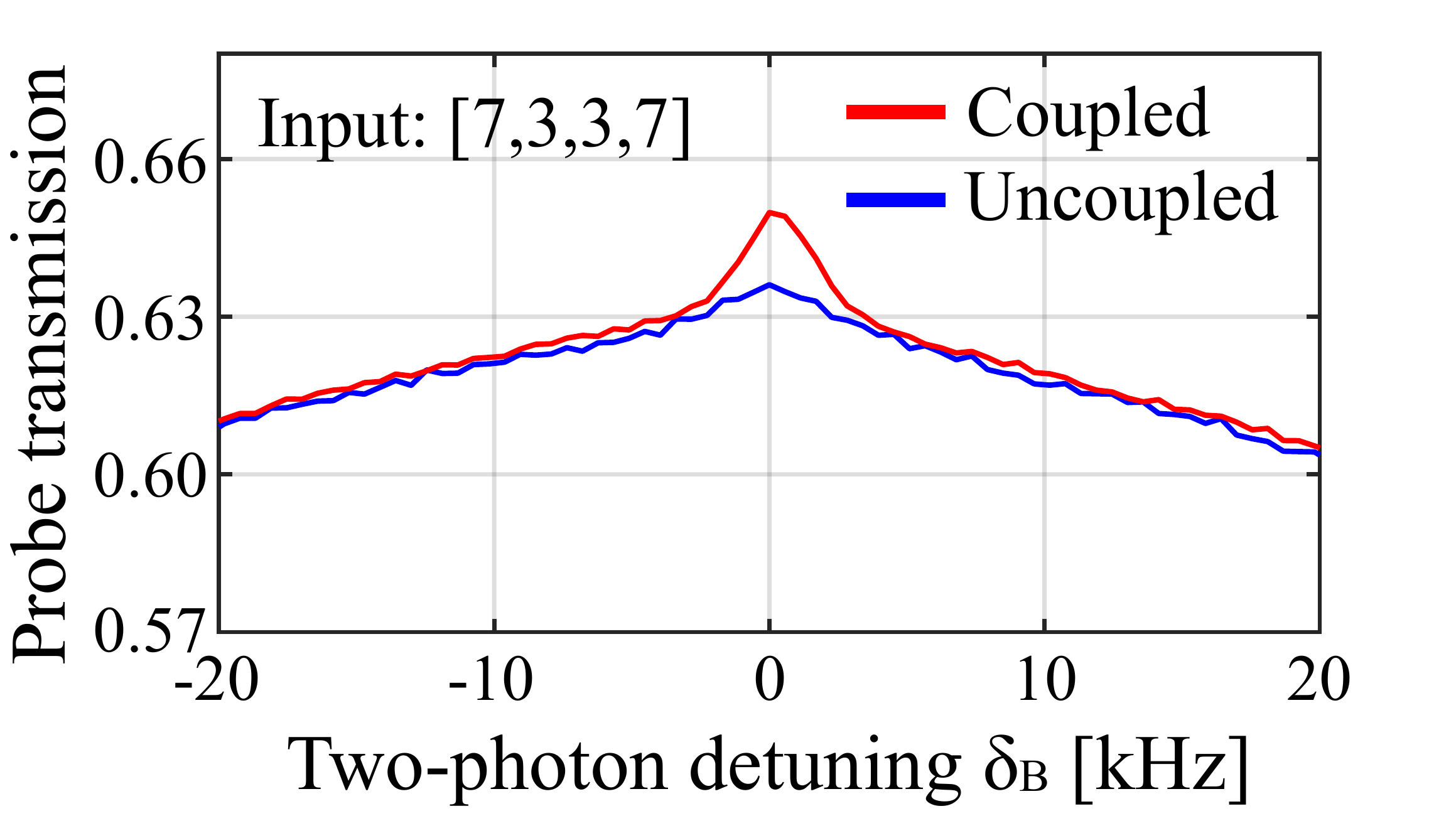}
	\caption{EIT spectra for a non-eigenstate input, showing the importance of the phase of the spin waves. Here, for coherence input of $[7,3,3,7]^T$, the coupled and uncoupled EIT has a pronounced difference, in contrast to the input eigenstate of $[-7,-3,3,7]^T$ which gives nearly identical coupled and uncoupled EIT spectra as shown in Fig.2 in the main text.}
\label{Fig. S3}
\end{figure}

In Fig.2 of the main text, we have presented the experimental and calculated EIT spectra corresponding to the four eigenstate inputs for the four-channel topological configuration, which shows good qualitative agreement. Here, as supplementary material, we emphasize that the phase or sign of the coherence $\rho_{12}$ is critical. For instance, as shown in Fig.~\ref{Fig. S3}, when the input is $[7, 3, 3, 7]^T$, the coupled and uncoupled EIT spectra are substantially different, as opposed to the case with $[-7, -3, 3, 7]^T$ input which has the same input laser power distribution but shows nearly identical coupled and uncoupled EIT spectra (see Fig.2 in the main text).

\begin{figure}
	\centering
	\includegraphics[width=0.9\textwidth]{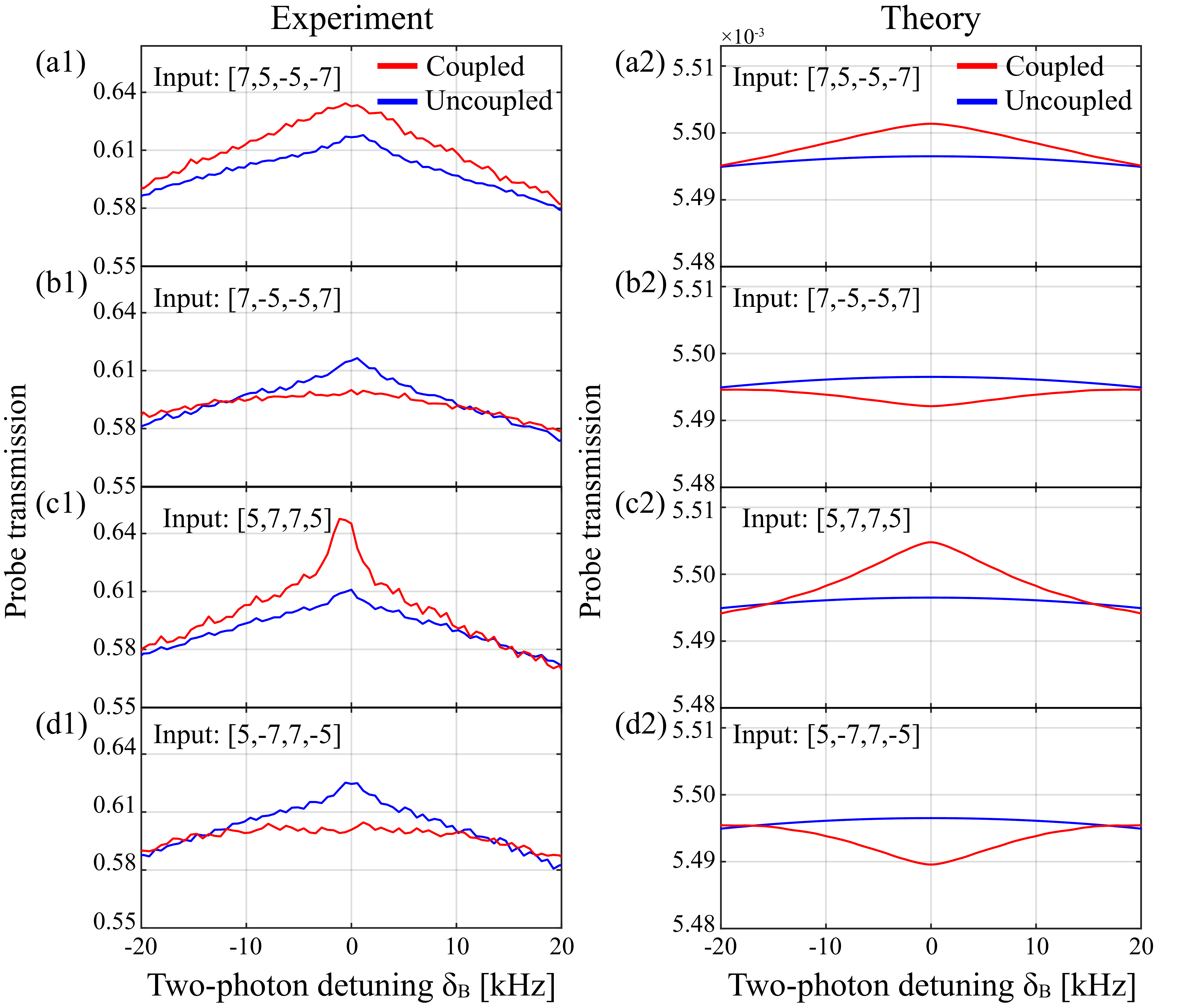}
	\caption{Measured and simulated EIT spectra for the four-channel non-topological configuration. The four rows correspond to the four eigenstate inputs respectively. Experimental spectra show the probe transmission normalized to far off resonant 100$\%$ transmission, and theory spectra display the calculated optical coherences.}
\label{Fig. S4}
\end{figure}

Furthermore, we show in Fig. \ref{Fig. S4} that for the four-channel non-topological configuration, qualitative agreement between experiment and theory can be also found. As expected, in contrast to the topological case, here the coupled and uncoupled EIT spectra are different for all the four eigenstate inputs, indicating the lacking of zero modes (with near-zero eigenvalue, i.e., coupled and uncoupled EIT are nearly identical).

\section{Experiment imperfections}

The EIT spectra of a single channel, or the EIT spectra in the uncoupled situation, displays a small unexpected spike on the top. This is likely due to the return of a small portion of atoms (upon bouncing off the cell wall, or with impurity background atoms in the cell) to the laser beams with remaining coherence. Previously, it has been found that in a buffer gas vapor cell or a wall-coated vapor cell, such return of atomic coherence plays a significant role in the atomic spectrum lineshape and gives rise to a very pointy EIT, as described by a Ramsey narrowing mechanism ~\cite{EIT-review2012}. Here, in the vacuum cell without buffer gas or wall-coating, the tiny pointy feature of the uncoupled EIT spectra indicates the existence of some residual all-to-all coupling, and is the major imperfection of our experiment because it resembles the pointy feature caused by inter-channel coupling of the coherence .

\begin{figure}
	\centering
	\includegraphics[width=0.5\textwidth]{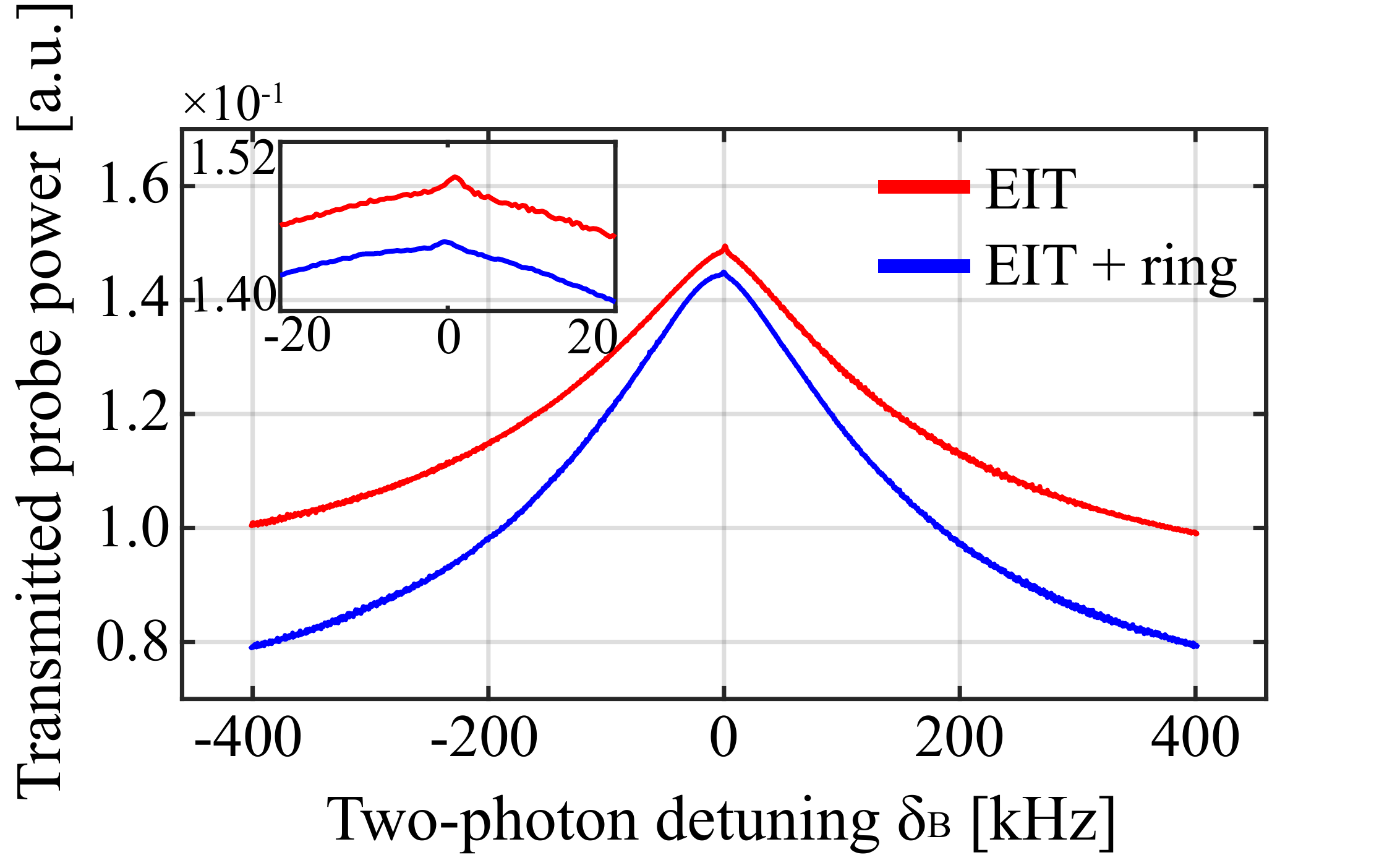}
	\caption{Experimental single-channel EIT spectra with and without a ring-shape optical pumping beam near the cell wall.}
\label{Fig. S5}
\end{figure}

To prove this is the case, experimentally, we use a ring shape laser beam near the cell wall to optically pump the atoms and destroy the coherence return from wall bouncing. Indeed, the pointy feature is slightly reduced as shown in Fig.~\ref{Fig. S5}. However, because of space constraints, this pumping beam cannot be kept on for the measurements of the topological features when four or six optical channels are turned on. In future experiments, vapor cells with different cross section geometry, rectangular or elliptical etc., can be used to alleviate the problem.

Such imperfection in the experiment affects the the positive maximal eigenvalue the most, as can be seen from Fig.3, because its eigenstate corresponds to same-sign coherence in all channels, which means the total coherence in the cell is maximal and so the returned coherence.

Another imperfection is that we still have some influence from the next nearest neighbor, because the $1/d$ scaling of the inter-channel coupling rate versus the channel spacing $d$ is not sharp enough. This accounts for the deviation of some of the measured eigenvalues from the prediction of our theoretical model which assumes only nearest neighbor coupling. For example, in the non-topological case, for the eigenstate $[-5,7,-7,5]^T$, let's consider the coherence change in Ch2 upon coupling: the two nearest neighbor contributes to coherence reduction due to the opposite sign in the coherence, while residual coupling with Ch4 gives coherence increase, which would decrease the amplitude of this eigenvalue. Indeed, as seen in Fig.3(b) of the main text, the first eigenvalue, corresponding to the eignestate input $[-5,7,-7,5]^T$, displays a smaller amplitude than the theoretical value. The eigenstate with same-sign coherence in all channels, corresponding to the 4th eigenvalues in Fig.3(a-b) is also affected by the next nearest neighbor coupling, rendering a noticeably larger eigenvalue than predicted.

\section{Monte Carlo simulation}

\begin{figure}
	\centering
	\includegraphics[width=0.87\textwidth]{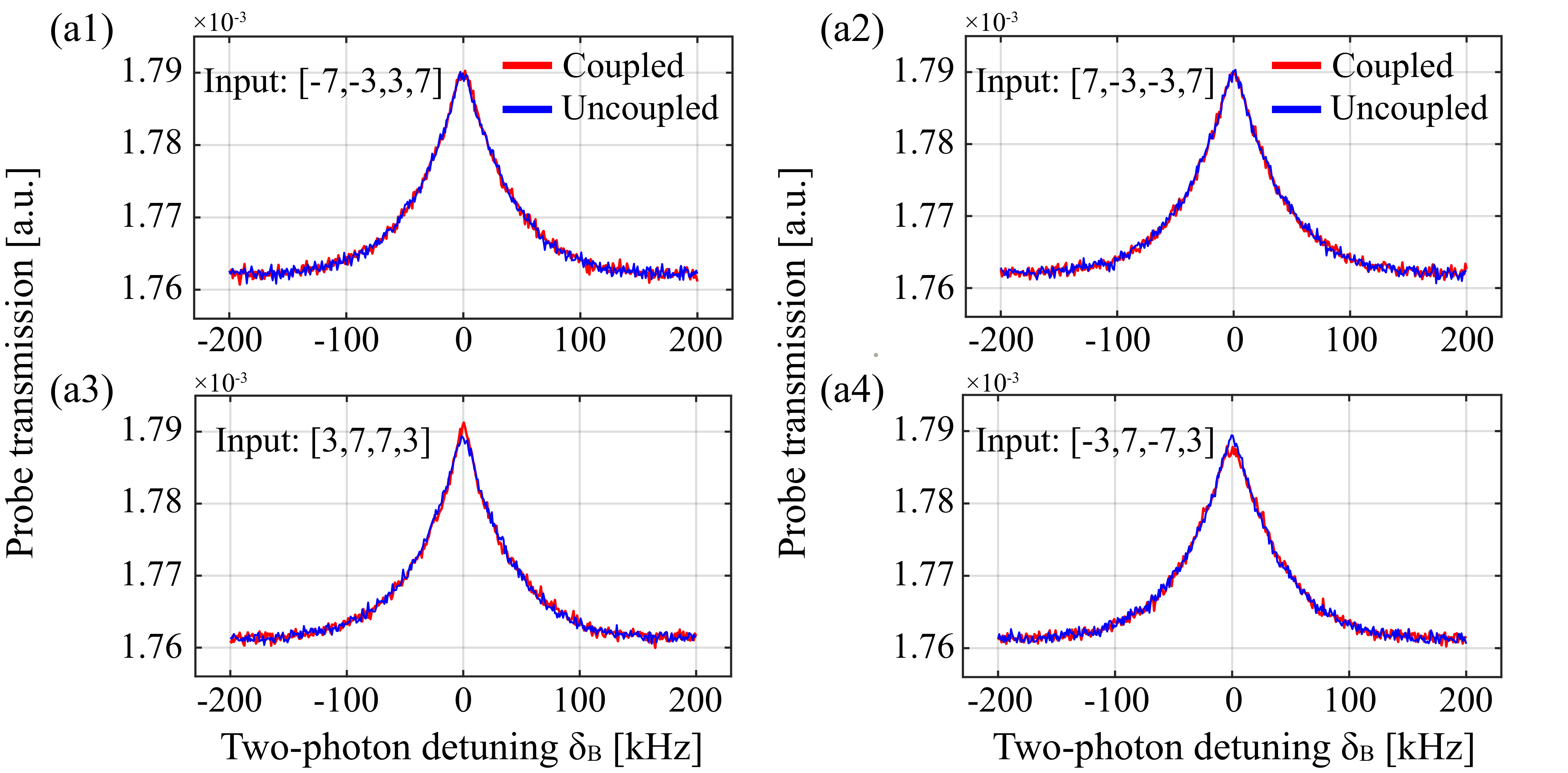}
	\caption{Monte Carlo simulation of uncoupled and coupled EIT spectra for the four-channel topological configuration. The four figures show the example EIT spectra (through the optical coherences) for the four eigenstates input, for the channels with relatively higher coherence, i.e., Ch1 in (a1), Ch1 in (a2), Ch2 in (a3), Ch2 in (a4). The spectra are in qualitative agreement with the experiment and the theoretical model described in the main text and in the first section here, with (a3) and (a4) indicating the zero modes which give nearly identical coupled and uncoupled EIT spectra. In the Monte Carlo simulation, flat-top laser beams with diameter of 1.2 mm are assumed, and the cell diameter is set to be 25 mm. The channel spacings are 3 mm and 6 mm, the same with the experiments. }
\label{Fig. S6}
\end{figure}

We have also implemented a two-dimensional Monte Carlo simulation for the vacuum cell experiment, to supplement the aforementioned theoretical model. The Monte Carlo model here is an extension of the one we developed previously~\cite{antiPT2016}, now adapted for the case of four channels and uncoated cell. Each time the atom bounces off the cell wall, we assume that the atom returns to thermal state with nearly no coherence. We let the atomic dynamics evolve until the system reaches its steady state, and then the physical quantities of interest are extracted. The results are averaged over 30 velocities satisfying the Maxwell-Boltzmann distribution, with each velocity averaging over 7000 different trajectories of the atomic movement. As in the experiment, the two-photon detuning is varied by changing the applied magnetic field. Fig.~\ref{Fig. S6} shows the calculated EIT shapes (displaying both the broad structure and the narrow structure caused by the SSH Hamiltonian) for the topological case, and the results agree qualitatively with the experiment and the simplified theoretical model. In particular, the two edge states (zero modes) shows nearly no difference between the coupled and uncoupled EIT spectra, indicating the near-zero eigenvalues which are observed in the experiment and the simplified theoretical model.

Finally, we note that the EIT lineshape from the Monte Carlo simulation better mimics the experimental EIT lineshapes than the simplified model, for instance, the uncoulped EIT is pointier than that of the Lorenzian lineshape obtained from the simplified theoretical model presented in the first section here. Such deviation from the Lorenzian lineshape results from the distribution of atom-light interaction time (transit time of the atoms through the laser beam) associated with the velocity distribution of the atoms.

\section{Generalized-SSH with ten lattice sites}

To construct a larger lattice under the constraints of the size of the cylindrical vapor cell and the clear aperture of the magnetic shield, we design the configuration with input laser beams arranged in an open-ring-like pattern containing ten optical channels in topological or trivial (non-topological) regime [shown in main text Fig.4]. Owing to the fact that the inter-channel coupling rate is proportional to $1/d$, and that there is nearly no ``blocking" effect for beyond-nearest-neighbor couplings as in the straight-line configuration, the next-nearest-neighbor (NNN) hopping terms in the present configuration should be considered with the NNN coupling rate $t_{A,B}$ approximately $\propto1/(d_{1} + d_{2})$. Furthermore, compared to the four-channel line-pattern, the ring-like pattern here has some additional experimental imperfections which result in slight deviation of the experimentally extracted eigenvalues from the theoretical ones: (a) Due to the much reduced distance between the optical channels and the cell wall, the local Markovian reservoir associated with the residual return of the ground state atomic coherence after wall collision causes a slight increase of the probe transmission, which leads to a small elevation of all the extracted eigenvalues. (b) In the open-ring-like configuration, there is unwanted coupling between the two end lattice sites, which has more influence on the extracted eigenstates of the edge states than on the bulk ones.

\begin{figure}
	\centering
	\includegraphics[width=0.5\textwidth]{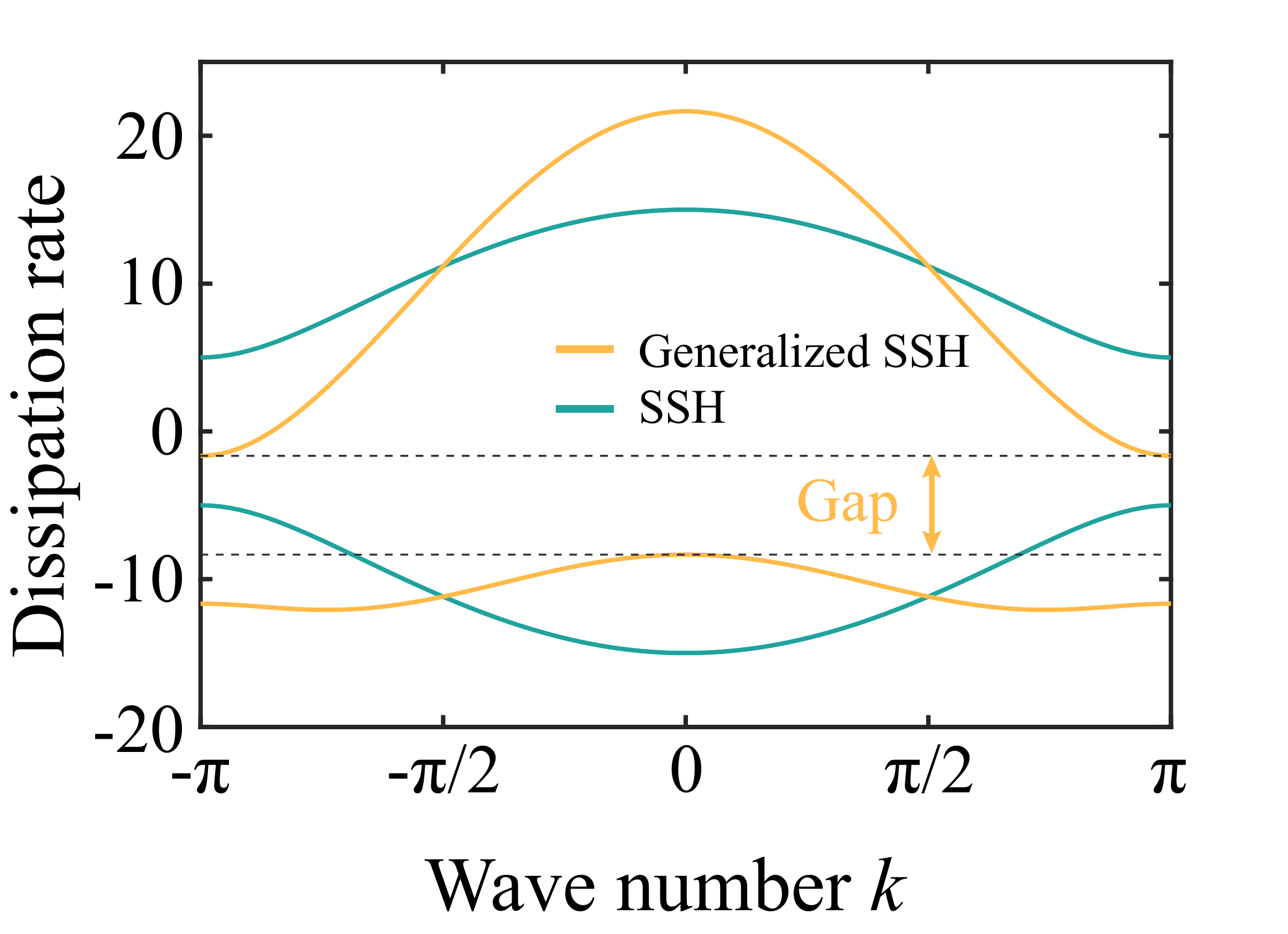}
	\caption{Calculated dissipation spectrum for the SSH and Generalized-SSH Hamiltonian in the momentum space. The gap for the Generalized-SSH model is indicated by the dashed lines. Parameters: $t$=3.33, $v$=5, $w$=10.}
	\label{Fig. S7}
\end{figure}
Theoretically, one can write down the Hamiltonian $\hat{H}$ describing the coupling between the coherence in the ten optical channels with NNN coupling:
\begin{equation}
	\label{s9}
	\tag{S9}
	\begin{aligned}
		\begin{split}
			\hat{H} =\ & ive^{i\delta_\textrm{B} d_{1}/\nu} \sum_{m} (\ket{a,m}\bra{b, m} + \textrm{H.C.})\\
			&+iwe^{i\delta_\textrm{B} d_{2}/\nu} \sum_{m} (\ket{b,m}\bra{a, m+1} + \textrm{H.C.})\\
			&+it_{\textrm{A}}e^{i\delta_\textrm{B} (d_{1}+d_{2})/\nu} \sum_{m} (\ket{a,m}\bra{a, m+1} + \textrm{H.C.})\\
			&+it_{\textrm{B}}e^{i\delta_\textrm{B} (d_{1}+d_{2})/\nu} \sum_{m} (\ket{b,m}\bra{b, m+1} + \textrm{H.C.})\\
		\end{split}
	\end{aligned}
\end{equation}
Here, $|t_{A}|$ and $|t_{B}|$ are the NNN hopping amplitudes between sublattices $A$ and $B$, respectively. We have set $|t_{A}|$=$|t_{B}|$=$t$ since the inter-channel coupling rate is proportional to $1/d$. While the atoms travel in sublattices $A$ or $B$, the ground state coherence will accumulate a phase  $\theta_{3}=\theta_{4}\approx\delta_B(d_{1}+d_{2})/\nu$.
Under periodic boundary conditions, we can make a Fourier transformation conveniently,
\begin{equation}
	\label{s10}
	\tag{S10}
	\begin{aligned}
		\begin{split}
			H(k)=
			\begin{pmatrix}
				i2t_{\textrm{A}}e^{i\theta_{3}}\cos{k} & ive^{i\theta_{1}}+iwe^{i(\theta_{2}-k)} \\
				ive^{i\theta_{1}}+iwe^{i(\theta_{2}+k)} & i2t_{\textrm{B}}e^{i\theta_{4}}\cos{k}
			\end{pmatrix}
		\end{split}
	\end{aligned}
\end{equation}
After diagonalizing the Hamiltonian, we get the eigenvalues,
\begin{equation}
	\label{s11}
	\tag{S11}
	\begin{aligned}
		\begin{split}
			E(k) = (it_{\textrm{A}}e^{i\theta_{3}} + it_{\textrm{B}}e^{i\theta_{4}}) cos(k) \pm \sqrt{\Delta(k)}
		\end{split}
	\end{aligned}
\end{equation}
where, $\Delta(k) = (it_{\textrm{A}}e^{i\theta_{3}} - it_{\textrm{B}}e^{i\theta_{4}})^2 cos^2(k) + (ive^{i\theta_{1}})^2 + (iwe^{i\theta_{2}})^2 + 2 (ive^{i\theta_{1}}) (iwe^{i\theta_{2}}) cos(k)$. We will focus on the experimental relevant situation where $\theta_{i}\approx0$ ($i=1,2,3,4$) and $t_A=t_B$. In this situation, we show in Fig.\ref{Fig. S7} the dissipation spectrum of the standard SSH model and Generalized-SSH model in momentum space, which indicates the broken of chiral symmetry due to the existence of NNN coupling.

\begin{figure}
	\centering
	\includegraphics[width=0.87\textwidth]{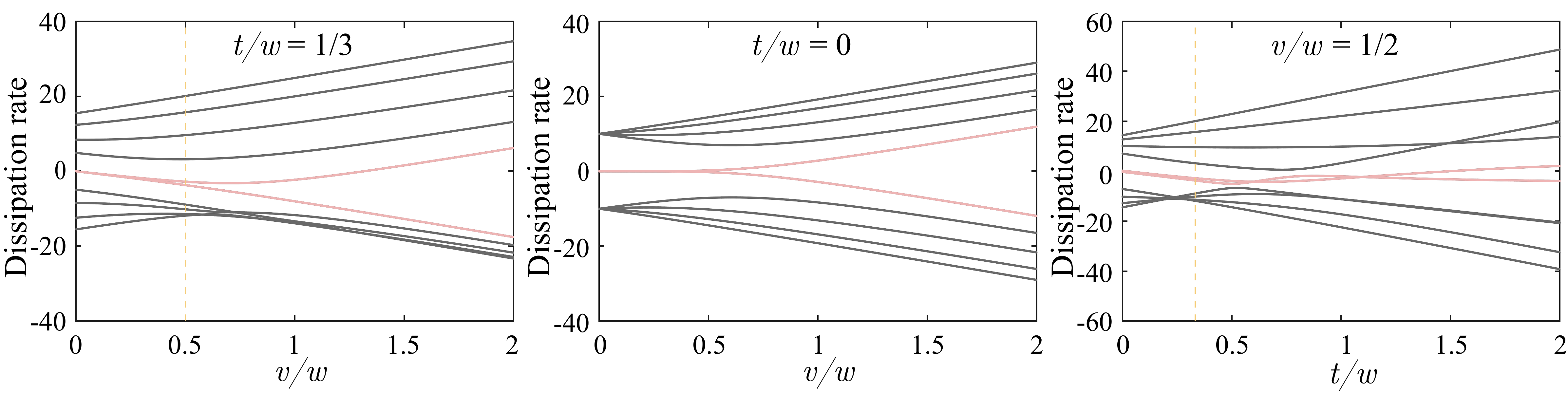}
	\caption{Calculated dissipation spectrum for the SSH model and Generalized-SSH model with 10 lattice sites, in real space. The vertical dashed lines (yellow) indicate our experiment parameter regime. The modes in pink are of particular interest (including the ``zero" modes). Parameter: $w=10$.}
	\label{Fig. S8}
\end{figure}

\begin{figure}
	\centering
	\includegraphics[width=0.5\textwidth]{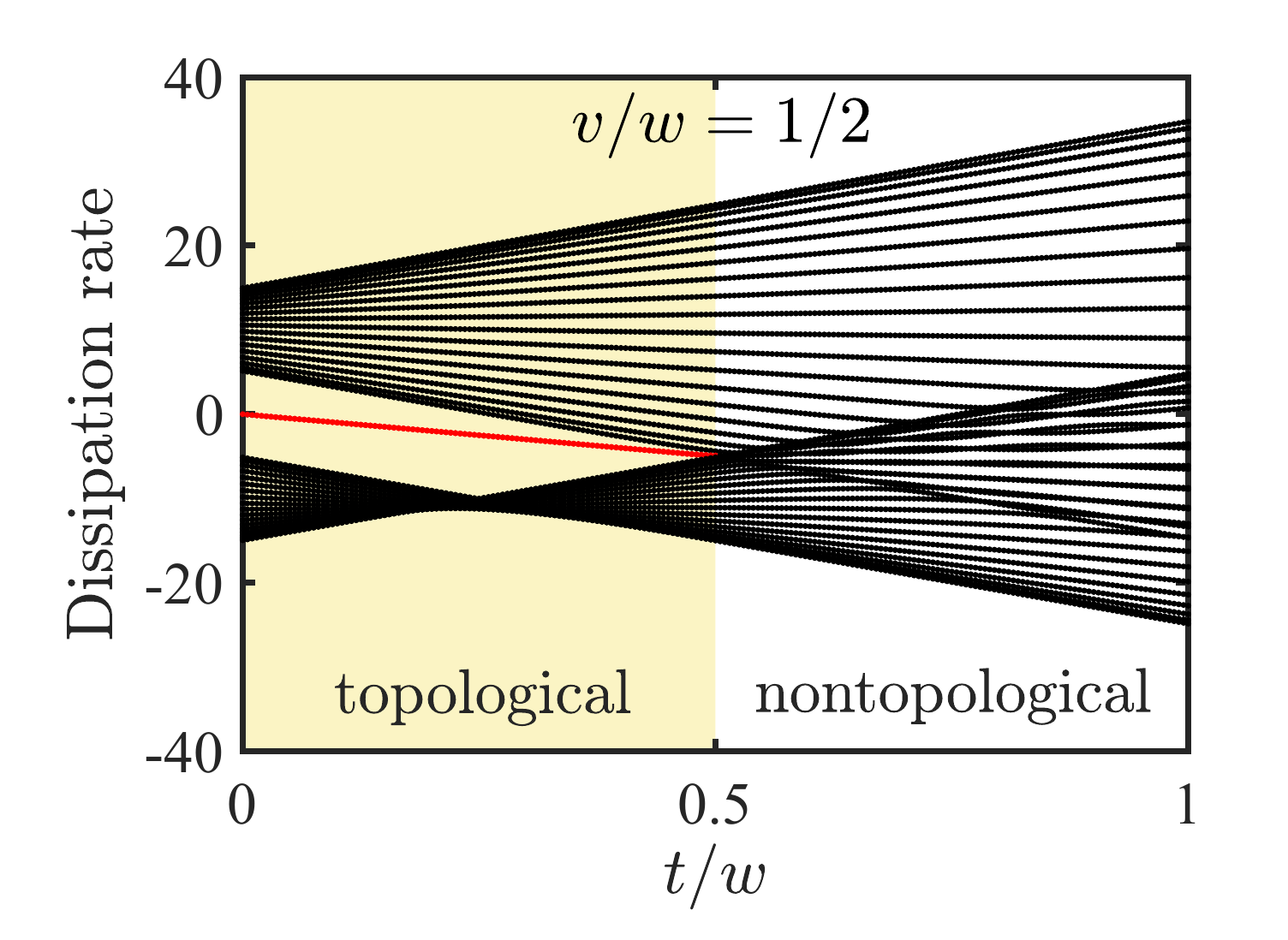}
	\caption{Calculated dissipation spectrum for the Generalized-SSH model with 40 lattice sites, in real space. The shaded region retains topology and the ``zero" modes (in red). Parameter: $w=10$.}
	\label{Fig. S9}
\end{figure}

The dissipative SSH model has chiral symmetry and inversion symmetry. The presence of NNN couplings with $t_A=t_B$ and $\theta_3=\theta_4=0$, breaks chiral symmetry, but preserves inversion symmetry.  However, by judiciously setting $v:w:t \approx \frac{1}{2}:1:\frac{1}{3}$ (see Fig.~\ref{Fig. S8}), we can make sure that the generalized-SSH Hamiltonian is still topological (see Fig.~\ref{Fig. S9} for a phase diagram), which means the existence of two (nearly degenerate) edge modes.

\bibliographystyle{apsrev}